\documentclass[aps,prb,twocolumn,floatfix,showpacs,citeautoscript,longbibliography,hyperlinks,superscriptaddress]{revtex4-1}

\usepackage{graphicx}
\usepackage{dcolumn}
\usepackage{bm}
\usepackage{bbold}
\usepackage{mathtools}
\usepackage{color}
\usepackage{transparent}
\usepackage[colorlinks=true,citecolor=blue]{hyperref}

\usepackage{etoolbox}
\usepackage{scalerel}
\usepackage{cleveref}
\usepackage{makecell}
\usepackage{braket}
\usepackage{float}
\usepackage[final]{changes}

\newcommand{\detector}[0]{\ensuremath{\mathrm{D}}}

\newcommand{\aalto}{QTF Centre of Excellence, Department of Applied Physics,\\ Aalto University, P.O. Box 13500, FI-00076, Aalto, Finland}
\newcommand{\unsw}{School of Electrical Engineering and Telecommunications,\\
University of New South Wales, Sydney, New South Wales 2052, Australia}

\newcommand{\npl}{National Physical Laboratory, Hampton Road, Teddington TW11 0LW, United Kingdom}
\newcommand{\strathclyde}{Department of Physics, SUPA, University of Strathclyde, Glasgow G4 0NG, United Kingdom}
\newcommand{\vtt}{VTT Technical Research Centre of Finland Ltd, P.O. Box 1000, FI-02044, VTT, Finland}

\begin{document}
\title{Waiting time distributions in a two-level fluctuator coupled to \\ a superconducting charge detector}

\author{M\'at\'e Jenei}
\thanks{These authors contributed equally to this work}

\affiliation{\aalto}
\author{Elina Potanina}
\thanks{These authors contributed equally to this work}
\affiliation{\aalto}
\author{Ruichen Zhao}
\thanks{ Current address: NIST 325 Broadway, Boulder, CO 80305, USA}
\affiliation{\unsw}
\author{Kuan Y. Tan}
\affiliation{\aalto}
\author{Alessandro Rossi}
\affiliation{\strathclyde}
\affiliation{\npl}
\author{Tuomo Tanttu}
\affiliation{\unsw}
\author{\\Kok W. Chan}
\affiliation{\unsw}
\author{Vasilii Sevriuk}
\affiliation{\aalto}

\author{Mikko M\"ott\"onen}
\affiliation{\aalto}
\affiliation{\vtt}
\author{Andrew Dzurak}
\affiliation{\unsw}
\begin{abstract}
We analyze charge fluctuations in a parasitic state strongly coupled to a superconducting Josephson-junction-based charge detector. The charge dynamics of the state resembles that of electron transport in a quantum dot with two charge states, and hence we refer to it as a two-level fluctuator. By constructing the distribution of waiting times from the measured detector signal and comparing it with a waiting time theory, we extract the electron in- and out-tunneling rates for the two-level fluctuator, which are severely asymmetric. 
\end{abstract}

{\let\newpage\relax\maketitle}

\section{Introduction}

Parasitic states including charge traps are present in almost all solid-state devices and there has been several proposals on how to avoid them. \cite{Holweg1992,Galperin1994,Galperin1995,Keijsers1996,Balkashin1998,Oh2006,Schriefl2006,Zimmerman2008,Burnett2014} Two-level fluctuators (TLFs), for example, substantially affect qubit coherence time \cite{Mottonen2006,Ku2005,Muller2009,Goetz2016} and degrade charge sensing\cite{Giblin2016}. However, if the time scales of charge fluctuations in a trap are significantly different from those of the operation of the actual device, their harmful effect can be mitigated. In silicon, TLFs have been characterized by various approaches using metallic single-electron transistors \cite{Zimmerman1997,Furlan2003,Sun2009,Pourkabirian2014,Zimmerman2008}, a scheme to which we contribute in this paper.

Electron waiting times have been investigated for a wide range of physical systems including quantum dots, \cite{Brandes2008,Welack2008,Welack2009,Welack2009Non,Thomas2013,Tang2014,Tang2014Full, Sothmann2014,Talbo2015,Rudge2016,Rudge20162,Ptaszynski2017,Rudge2018,Stegmann2018,Kleinherbers2018,Tang2018,Rudge2019} coherent conductors, \cite{Albert2012,Haack2014} molecular junctions \cite{SeoaneSouto2015, Kosov2018}, and superconducting systems. \cite{Rajabi2013,Dambach2015,Dambach2016,Albert2016,Chevallier2016, Walldorf2018, Mi2018} Distributions of waiting times contain complementary information on charge transport properties which is not necessarily encoded in the full counting statistics (FCS) and vice versa \cite{Brandes2008}. For example, waiting-time distributions capture the interference effects in double-dot setups \cite{Welack2008}, reveal the correlations in multichannel systems \cite{Tang2014, Dasenbrook2015}, allow to separate slow and fast dynamics in Cooper-pair splitters \cite{Walldorf2018}, resolve few-photon processes \cite{Brange2019}, and even investigate the topological superconductivity in hybrid junctions \cite{Mi2018}. In dynamic, periodically driven systems, waiting-time distributions are clear indicators of regular single-electron transport \cite{Albert2011,Dasenbrook2014,Potanina2017,Burset2019}. Furthermore, waiting-time distributions were used in a recent experiment \cite{Gorman2017} to optimize single-electron spin-readout fidelity. 

\begin{figure}
\centering
\includegraphics[width=\linewidth]{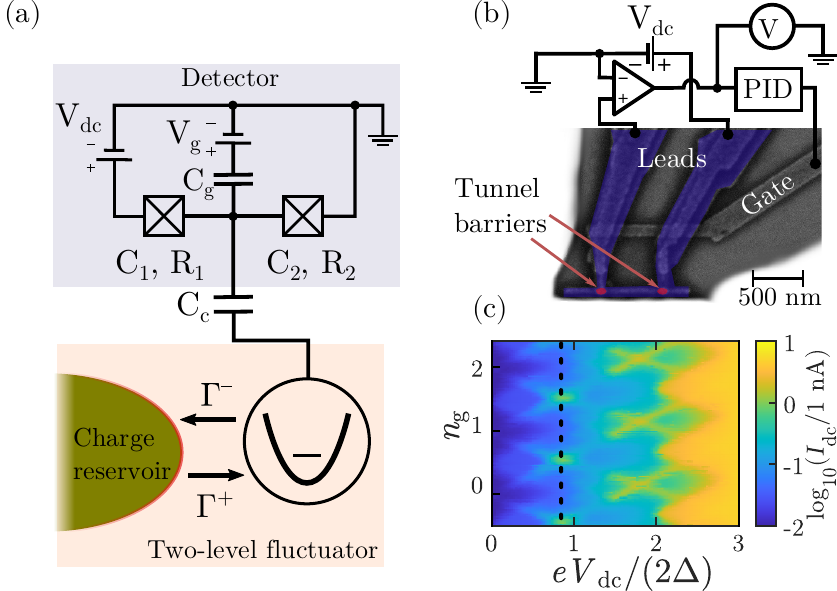}
\caption{(a) Schematic diagram of the investigated system which consist of a Josephson-junction-based detector (top) and a two-level system (bottom) that is in tunnel contact with a charge reservoir. All used symbols are described in the main text. (b) False-color scanning electron micrograph of the charge detector similar to that in the measurements together with a circuit diagram of the experimental setup. (c) Detector current as a function of the normalized bias voltage $eV_{\rm{dc}}/(2\Delta)$ and the gate charge $n_{\rm{g}} = V_{\rm{g}}C_{\rm{g}}/e$,  where $\Delta$ is the superconducting gap. The dashed line indicates the bias point for the measurement. The parameters of the measured device are $\Delta = 195 ~\mu$eV, $R_{\rm{T}} = R_1 + R_2= 180 ~$k$\Omega$, and \mbox{$E_{\rm{ch}} = 240 ~\mu$eV.}}
\label{fig:figure1}
\end{figure}

In this work, we investigate electronic waiting times between charge transitions in and out of a parasitic state to directly extract the time-scales of such TLF. We employ a superconducting single-electron transistor (SSET) to monitor the switching events on the TLF and apply a continuous electrostatic feedback on the detector to maintain a constant charge sensitivity, which tends to fasten an asymmetry in the detected in- and out-tunneling rates. 

This paper is organized as follows: In Sec.~\ref{sec:WTD}, we provide a short overview of the waiting time theory \cite{Brandes2008} and illustrate  these  concepts  by  evaluating  the distribution of electron waiting times for a TLF in Secs.~\ref{sec:TLF} and ~\ref{sec:TLFandDetector}. We treat the two-level fluctuator as a potential well that is in tunnel contact with a charge reservoir as illustrated in in \mbox{Fig.~\ref{fig:figure1}(a).} In Sec.~\ref{sec:Experiment}, we discuss our experimental setup. In Sec.~\ref{sec:Results} we present our measurement results and compare them with the waiting-time-distribution theory. Since our model assumes sequential in- and out-tunneling, we calculate the FCS of the switching events and compare the results with the waiting times. Finally, in Sec.~\ref{sec:Conclusions}, we conclude our work.  

\section{Electron waiting times}
\label{sec:WTD}

The time that passes between two subsequent single-electron tunneling events of the same type is usually referred to as the electron waiting time $\tau$.\cite{Brandes2008,Albert2011,Albert2012} The single-electron tunneling process has a stochastic nature and therefore is described by a waiting-time distribution (WTD) function $\mathcal{W}(\tau)$. For stationary Markovian transport problems, the WTD relates to the idle-time probability $\Pi(\tau)$ as \cite{Albert2012,Haack2014}
\begin{equation}
    \mathcal{W} (\tau) = \langle \tau \rangle \partial_{\tau}^2 \Pi(\tau),
\label{eq:WTD}
\end{equation}
where $\Pi(\tau)$ is the probability of having no tunneling events during a time span $\tau$. The mean waiting time $\langle \tau \rangle$ can be expressed in terms of the idle-time probability as\cite{Albert2012,Haack2014} $\langle \tau \rangle = \int_0^{\infty} d \tau  \mathcal{W}(\tau)\tau=-1/\dot{\Pi}(\tau=0)$.

The statistics of single-electron tunneling events is captured by the probability $P(n,t)$ of having $n$ tunneling events of the chosen type during the time span $[t_0,t_0+t]$.\cite{Bagrets2003,Flindt2008,Flindt2010} However, we only need to know the idle-time probability $\Pi(\tau) = P(n=0,\tau)$ to obtain the WTD. 
In FCS, the moment generating function
\begin{equation}
    \mathcal{M}(\chi,t) = \sum\limits_{n=0}^{\infty} P(n,t) e^{in\chi},
\label{eq:MGF}
\end{equation}
provides us with all the moments of $n$ as $\langle n^m\rangle(t) = \partial_{i\chi}^m \mathcal{M}(\chi,t)\vert_{i\chi\rightarrow0}$. From Eq.~\eqref{eq:MGF} we observe that $\mathcal{M}(i\infty,t)=P(n=0,t)$ is exactly the idle-time probability. Next, we utilize these concepts by evaluating the WTDs for a two-level fluctuator.

\subsection{Waiting times in a two-level fluctuator}
\label{sec:TLF}

We describe the parasitic state as a single-electron box consisting of a nano-scale island, the charge dynamics of which is governed by the master equation
\begin{equation}
      \frac{\textrm{d}}{\textrm{d}t} \textrm{p}(t)  = \mathbf{L} \textrm{p}(t) ,
\label{eq:RateEq}
\end{equation}
where the vector $ \textrm{p}(t)=[p_0(t),p_1(t)]^{T}$ contains the probabilities $p_0(t)$ and $p_1(t)$ for the island to be empty or occupied by $1$ electron, respectively, and the rate matrix $\mathbf{L}$ describes the transitions between $0$ and $1$ charge states of the island.

 We partition the rate matrix as $\mathbf{L}=\mathbf{L}_0+\mathbf{J}_++\mathbf{J}_{-}$ with jump operators $\mathbf{J}_{\pm}$ describing charge transfers to and from the island, respectively \cite{Benito2016}. We resolve the probability vector $\textrm{p}(n,t)$ such that it accounts for the number of tunneling events $n$. The $n$-resolved equations of motion, $\frac{\textrm{d}}{\textrm{d}t}\textrm{p}(n,t) =\mathbf{L}_0 \textrm{p}(n,t) +\mathbf{J}_+\textrm{p}(n-1,t) +\mathbf{J}_{-}\textrm{p}(n+1,t)$, are decoupled by introducing the counting field $\chi$ via the definition $\textrm{p}(\chi,t)\equiv\sum_n\textrm{p}(n,t) e^{in\chi}$. We then arrive at a modified master equation for $\textrm{p}(\chi,t)$ 
\begin{equation}
      \frac{\textrm{d}}{\textrm{d}t}\textrm{p}(\chi,t) = \mathbf{L}(\chi)\textrm{p}(\chi,t).
\label{eq:RateEq_chi}
\end{equation}
For $\chi=0$ in Eq.~\eqref{eq:RateEq_chi}, we recover the original master equation \eqref{eq:RateEq}. Further on, we focus on the waiting times between the into-the-island tunneling events, and hence set $\mathbf{J}_{+}=0$. In this case, the modified rate matrix $\mathbf{L} (\chi)$ assumes the form
\begin{equation} \label{eq:RateMatrix}
\mathbf{L}(\chi) = \left( \begin{array}{cc}
     -\Gamma^{+} &  \Gamma^{-} \\\\
     e^{i \chi} \Gamma^{+} & -\Gamma^{-}
\end{array}  
\right).
\end{equation}
We have included the counting factor $e^{i \chi}$ in the lower off-diagonal element together with $\Gamma^{+}$, corresponding to counting the number of tunneling events into the parasitic state.\cite{Bagrets2003,Flindt2008,Flindt2010}
The solution of the modified master equation formally reads:
\begin{equation}
    \textrm{p}(\chi,t)= e^{\mathbf{L}(\chi) t} \textrm{p}(\chi,t).
\end{equation}
The idle-time probability follows as $\Pi (\tau) = \sum_j p_j(i \infty,t)$, where $p_j(i \infty,t)$ is an $j$th component of the vector $\textrm{p}(\chi,t)\vert_{i\chi \rightarrow i \infty}$. For a given rate matrix \eqref{eq:RateMatrix} the solution is analytic and we are  able to evaluate the WTD for a two-level fluctuator using relation \eqref{eq:WTD} as
\begin{equation}
\mathcal{W}(\tau) = \Gamma^{+}\Gamma^{-}\frac{e^{-\Gamma^{-}\tau}-e^{-\Gamma^{+}\tau}}{\Gamma^{+}-\Gamma^{-}},
\label{eq:waiting_time_distr}
\end{equation}
where $\tau$ is the waiting time between single-electron subsequent tunneling events from the charge reservoir to the parasitic state. We continue with another example of WTD, where we take into account the effect of finite detection time.

\subsection{Residence times in a two-level fluctuator coupled to a detector}
\label{sec:TLFandDetector}

\begin{figure*}
\centering
\includegraphics[width=\linewidth]{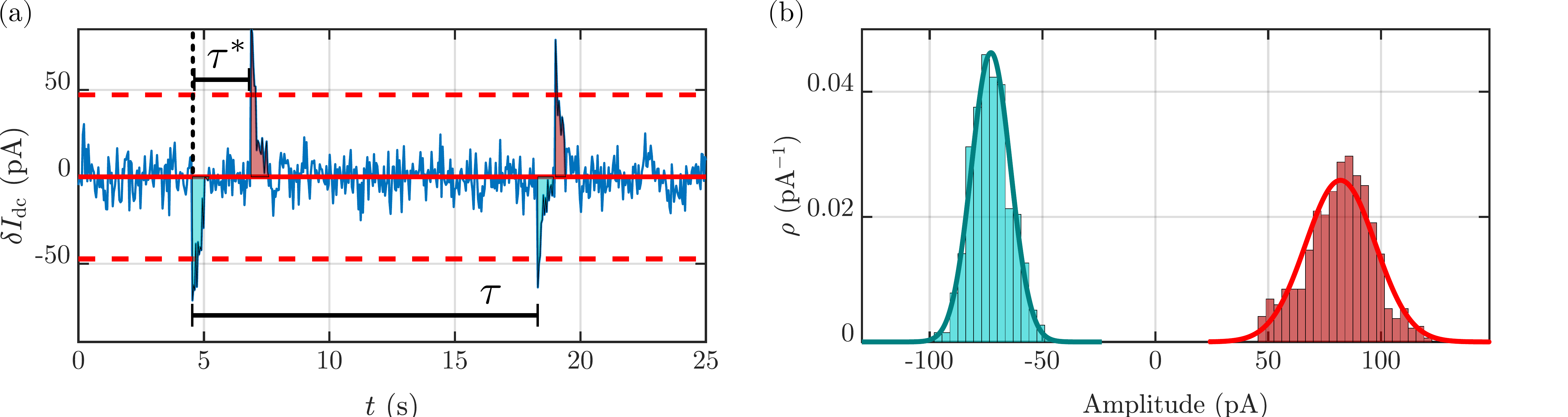}
\caption{
(a) Electric current offset of the detector as a function of time (blue line). The two characteristic parameters, the waiting time ($\tau$) and the residence time ($\tau^*$) are indicated. The red dashed lines correspond to $\pm$5$\sigma$ white-noise levels and the red solid line depicts the PID set current, which is at the positive slope on the dashed line in Fig.~\ref{fig:figure1}(c). The shaded areas indicate the signal corresponding to the jump events. (b) Probability density of the charge signal, $\rho$, obtained by monitoring the jump amplitudes. The colors match with those of panel (a). }
\label{fig:figure2}
\end{figure*}

Let us focus on the time that the electron spends in the parasitic state known as the residence time $\tau^*$ \cite{Talkner2005}. We use the term residence time to avoid confusion with the waiting time discussed in the previous section. When the residence time becomes comparable to the inverse of the detector bandwidth, $ 2\pi/ \Gamma^{\detector}$, we have to take it into account when evaluating the distribution of residence times.

Due to the finite bandwidth of the detector, we may fail to detect some of the tunneling events. Therefore, we add two more possible states when formulating the rate equation.\cite{Naaman2006} The revised probability vector reads $\textrm{p}^*(t)=[p_{00}(t),p_{10}(t),p_{01}(t),p_{11}(t)]^{T}$. Probabilities $p_{01}(t)$ and $p_{10}(t)$ correspond to the situation when parasitic state is empty or occupied,  respectively, but it has not been detected. The detected empty and occupied states are denoted by $p_{00}(t)$ and $p_{11}(t)$, respectively.

While actual charge transitions between the occupied state $p_{11}(t)$ to the undetected empty state $p_{01}(t)$ may occur several times before the empty state $p_{00}(t)$ gets detected, we obtain the experimentally observed residence time distribution by counting the events $p_{01}(t) \rightarrow p_{00}(t)$.
After including the effect of the finite detector bandwidth $\Gamma^{\detector}/(2\pi)$ in the rate equation, we arrive at the following rate matrix \cite{Gustavsson2007,Flindt2009}:

\begin{equation} \label{eq:RateMatrixD}
\mathbf{L}(\chi) = \left( \begin{array}{cccc}
     -\Gamma^{+} & \Gamma^{-} & \Gamma^{\detector}e^{i \chi} & 0  \\\\
       \Gamma^{+} & -(\Gamma^{-}+\Gamma^{\detector}) & 0 & 0  \\\\
       0 & 0 & -(\Gamma^{+}+\Gamma^{\detector}) & \Gamma^{-}  \\\\
         0 & \Gamma^{\detector} & \Gamma^{+} & -\Gamma^{-}  
\end{array}  
\right).
\end{equation}
In analogy to the previous section, we solve the modified rate equation given the rate matrix Eq.~\eqref{eq:RateMatrixD} and obtain the idle-time probability $\Pi(\tau^*)$. We evaluate the detected residence time distribution using double differentiation as $\mathcal{W}_r(\tau^*)=\langle \tau^* \rangle \partial^2_{\tau^*}\Pi(\tau^*)$ and obtain

\begin{equation}
\label{eq:residencetime_distr}
\mathcal{W}_r(\tau^*) = \frac{ \Gamma^{\detector}\Gamma^{-}}{\Lambda}e^{-\tau^*\sum_{\alpha}\Gamma^{\alpha}/2} \sinh[\tau^*\Lambda/2],
\end{equation}
where $\Lambda = \sqrt{(\sum_{\alpha}\Gamma^{\alpha})^2-4\Gamma^{\detector} \Gamma^-}$ and $\alpha = +, -, \rm{D}$. If we assume the detector to be perfect $\Gamma_D \rightarrow \infty$, we recover the exponential decay of the occupied parasitic state $\mathcal{W}_{\infty}(\tau^*) = \Gamma^-e^{-\tau^*\Gamma^-}$.

\section{Experimental setup}
\label{sec:Experiment}
We employ a superconducting charge detector which consists of two Al/Al$_2$O$_3$/Al Josephson junctions in series with resistances and capacitances $R_i$, $C_i$, $i=1,2$ and a superconducting gap $\Delta$, which form a charge island with a charging energy of $E_{\rm{ch}}=e^2/C_{\Sigma}$, where $C_{\Sigma}=C_1+C_2+C_{\rm{c}}+C_{\rm{g}}$, $e$ is the elementary charge, $C_{\rm{g}}$ is the capacitive coupling between the gate and the island, and $C_{\rm{c}}$ is the mutual capacitance between the TLF and the detector island. The device is fabricated on a \mbox{500-$\mu$m-thick} high-resistivity silicon wafer with 5-nm-thick high{-}purity field oxide. The measurements are carried out in a cryostat that has a base temperature of $T = 100 $ mK. Figure \ref{fig:figure1}(b) shows the top view of a detector that is nominally identical to the device that is used to carry out the presented experiments. The Al leads provide galvanic contacts between the tunnel junctions and the gate voltage $V_{\rm{g}}$ is used to tune the charge sensitivity of the SSET. The bias voltage $V_{\rm{dc}}$ controls the electrochemical potential of the lead, whereas the other lead is connected to a room temperature transimpedance amplifier. At the output port of the amplifier, we measure the voltage and transmit the signal to a proportional-integral-derivative (PID) feedback circuit that maintains the gate charge point, and hence the charge sensitivity of the detector. 

The charge stability of the detector, shown in \mbox{Fig.~ \ref{fig:figure1}(c),} reveals how to associate the measured dc current
with the gate-tunable charge point of the superconducting island. To minimize the detector backaction on the two-level fluctuator, we choose the bias voltage that corresponds to the double-Josephson-quasiparticle process\cite{Clerk2002,Xue2009}. 

\section{Results}
\label{sec:Results}
The signal produced by the TLF is read out using the detector such that the PID controller is utilized and tuned to a setpoint near the highest sensitivity. In addition to the white noise, we observe a systematic signal, as exemplified in \mbox{Fig.~\ref{fig:figure2}(a)}. The jumps are distinguished from the white noise when the amplitude of the detector signal exceeds the 5$\sigma$ white-noise level. Furthermore, the detector signal is filtered with a two-point moving averaging window, which results in a detector bandwidth of \mbox{$\Gamma^{\rm{D}}/(2\pi) = 11.25$ Hz}.

 First, a jump with a negative sign appears and the detector current offset takes a negative value. Then, the PID steers back to the setpoint using the detector gate. After $\tau^*$ residence time, another jump appears, but always with the opposite, positive sign compared to the first jump. The consecutive jump occurs at $\tau$ counted from the previous negative jump. The probability distribution of the jump amplitudes, shown in \mbox{Fig.~\ref{fig:figure2}(b)}, seems Gaussian, however the mean values and variances are different for the different jump directions.

\begin{figure}
\centering
\includegraphics[width=0.47\textwidth]{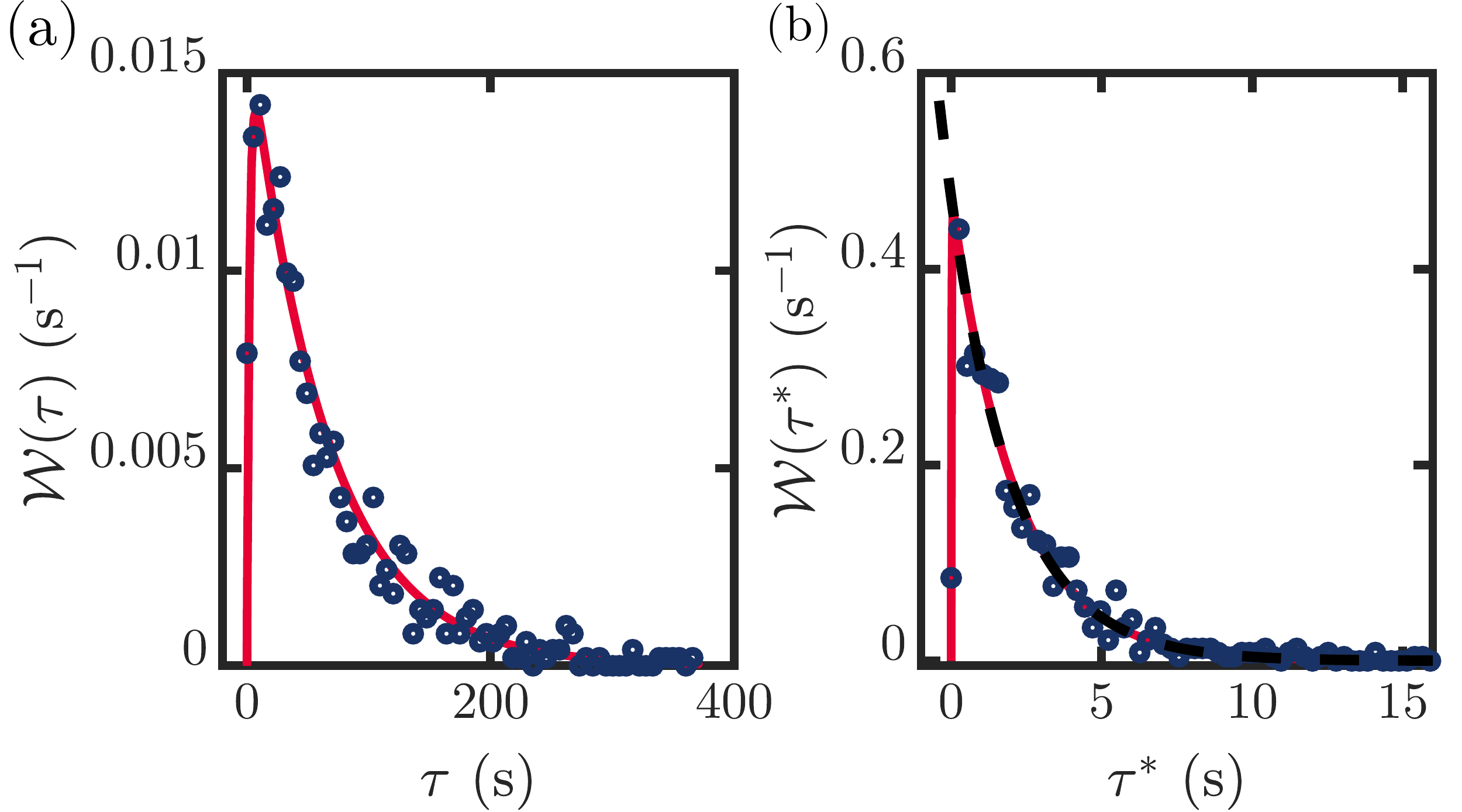}
\caption{(a) Measured (circles) and fitted (line) waiting time distribution of the charge jumps according to Eq.~\eqref{eq:waiting_time_distr}. (b) Measured (circles) and fitted residence time distribution of the charge jumps in the case of finite detector bandwidth (line) and ideal detector (dashed line) according to Eq.~\eqref{eq:residencetime_distr}. The data has been recorded as illustrated in \mbox{Fig.~\ref{fig:figure2}(a).}}
\label{fig:figure3}
\end{figure}

Since the switching between the two states is associated with single-electron transitions, $\tau$ and $\tau^*$ can be extracted from traces such as that shown in \mbox{Fig.~\ref{fig:figure2}(a)}. The total duration of the measured time trace is 16 h that provides approximately 1000 back-and-forth tunneling events, giving rise to the WTD shown in \mbox{Fig.~\ref{fig:figure3}(a)}. 

We obtain the parameters $\Gamma^+$ and $\Gamma^-$ by fitting \mbox{Eq.~\eqref{eq:waiting_time_distr}} to the data in \mbox{Fig.~\ref{fig:figure3}(a)}. The fitted waiting-time distribution, where the tunneling rates are \mbox{$\Gamma^+ = 15.8 \times 10^{-3}~$s$^{-1}$} and \mbox{$\Gamma^- =  473.2\times10^{-3}~$s$^{-1}$}, agrees well with the measured distribution. The significant difference between $\Gamma^+$ and $\Gamma^-$ originate from the energy splitting of the two charge states and from the feedback; for every switching event the detector gate induces a compensating electrostatic field that not only affects the operation of the SSET, but also tends to polarize the TLF towards the opposite state. Since the average residence time  $\langle\tau^*\rangle$ is significantly shorter than $\langle\tau\rangle$ and comparable to  $1/\Gamma^{\rm{D}}$, we fit Eq.~\eqref{eq:residencetime_distr} to the residence time distribution, shown in \mbox{Fig.~\ref{fig:figure3}(b)}, by first fixing $\Gamma^+$ and $\Gamma^-$ according to the WTD. Thus, the only fitting parameter in the model is $\Gamma^{\rm{D}}$, with the estimated $\Gamma^{\rm{D}}/(2\pi) = 10.4~$Hz based on the fit. The fitted $\langle \Gamma^{\rm{D}} \rangle$ deviates roughly 7.6\% from the nominal detector bandwidth. 

\begin{figure}
\centering
\includegraphics[width=0.47\textwidth]{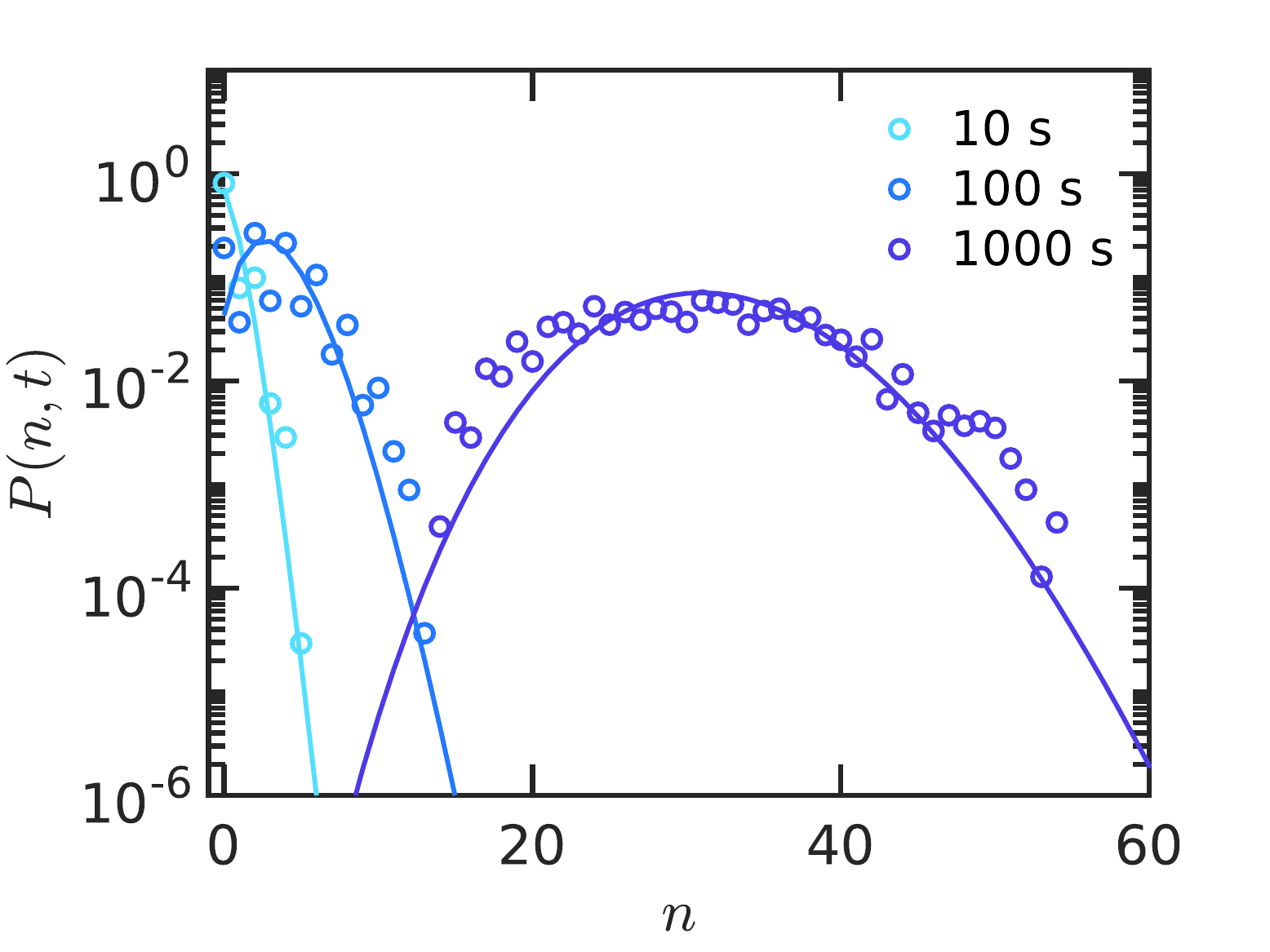}
\caption{Measured (circles) and fitted (lines) FCS of sequential single-electron tunneling events for three different time windows. }
\label{fig:FCS}
\end{figure}

From the measured state transitions we can also construct the counting statistics $P(n,t)$ used in Eq.~\eqref{eq:MGF}, where $n$ denotes the number of switching events regardless of the direction. The limited number of total jumps requires a moving time window to achieve the FCS with different of the time window lengths. Therefore,  consecutive time windows are shifted by 1 s and used in the average over the \mbox{16-h} time trace. The Markovian single-electron tunnelling in a dot coupled to a charge reservoir follows the Poisson distribution\cite{Maisi2014}. If we assume an effective tunneling rate $\Gamma^{\rm{eff}}$, we can fit a Poisson distribution to the data shown in Fig.~\ref{fig:FCS}. The extracted effective tunneling rate is $\Gamma^{\rm{eff}} = 31.5 \times10^{-3}~$s$^{-1}$, which can be explained by the branching of jumps observed in \mbox{Fig.~\ref{fig:figure2}(a)}. More precisely, we effectively observe two events with a rate of $\Gamma^{+}$, since after each in-tunneling event there is an out-tunneling event with a brief delay.

\section{Conclusions}
\label{sec:Conclusions}
We have demonstrated that WTD is useful in extracting the time scales of a TLF. Including the finite bandwidth of the detector in the waiting-time theory provides us with an accurate distribution of the waiting times in the charge trap. In contrast to WTD, full counting statistics covers long time scales and delivers mean values, noise, and high-order moments. Both approaches are powerful tools and they are connected \cite{Brandes2008}. From the observed probability distribution $P(n,t)$, we conclude that charge transitions in the TLF constitute a Poisson process. The distribution of waiting times, Eq.~\eqref{eq:waiting_time_distr}, indicates that there are two Poisson processes with different rates $\Gamma^+$ and $\Gamma^-$ in the two-level fluctuator. It requires less steps to evaluate the WTD than to evaluate the full counting statistics $P(n,t)$. Moreover, the waiting-time formalism allows to evaluate the WTD analytically where as an analytic expression for $P(n,t)$ is not currently available.
Waiting times belong to the short-time-scale statistics and are sensitive to the finite detector bandwidth. Here, we have shown how the distribution of waiting times has obvious advantages in extracting the short time scales of the system.

\begin{acknowledgments}
E.P. thanks C. Flindt for helpful comments.
This work was carried out as part of the Academy of Finland Centre of Excellence program 312300 and grants 308161, 314302, and 316551.  E.P. acknowledges support from the Vilho, Yrj\"o  and  Kalle  Foundation  of  the  Finnish  Academy of  Science  and  Letters  through  the  grant  for  doctoral studies. We acknowledge the provision of facilities and technical support by Aalto University at OtaNano - Micronova and the NSW Node of the Australian National Fabrication Facility, where the devices were fabricated. We acknowledge funding from the Joint Research Project ‘e-SI-Amp’ (15SIB08) and the Australian Research Council (DP160104923). This project has received funding from the European Metrology Programme for Innovation and Research (EMPIR) co-financed by the Participating States and from the European Union Horizon 2020 research and innovation programme.   
\end{acknowledgments}

\bibliography{My_Collection}

\begin{thebibliography}{64}%
\makeatletter
\providecommand \@ifxundefined [1]{%
 \@ifx{#1\undefined}
}%
\providecommand \@ifnum [1]{%
 \ifnum #1\expandafter \@firstoftwo
 \else \expandafter \@secondoftwo
 \fi
}%
\providecommand \@ifx [1]{%
 \ifx #1\expandafter \@firstoftwo
 \else \expandafter \@secondoftwo
 \fi
}%
\providecommand \natexlab [1]{#1}%
\providecommand \enquote  [1]{``#1''}%
\providecommand \bibnamefont  [1]{#1}%
\providecommand \bibfnamefont [1]{#1}%
\providecommand \citenamefont [1]{#1}%
\providecommand \href@noop [0]{\@secondoftwo}%
\providecommand \href [0]{\begingroup \@sanitize@url \@href}%
\providecommand \@href[1]{\@@startlink{#1}\@@href}%
\providecommand \@@href[1]{\endgroup#1\@@endlink}%
\providecommand \@sanitize@url [0]{\catcode `\\12\catcode `\$12\catcode
  `\&12\catcode `\#12\catcode `\^12\catcode `\_12\catcode `\%12\relax}%
\providecommand \@@startlink[1]{}%
\providecommand \@@endlink[0]{}%
\providecommand \url  [0]{\begingroup\@sanitize@url \@url }%
\providecommand \@url [1]{\endgroup\@href {#1}{\urlprefix }}%
\providecommand \urlprefix  [0]{URL }%
\providecommand \Eprint [0]{\href }%
\providecommand \doibase [0]{http://dx.doi.org/}%
\providecommand \selectlanguage [0]{\@gobble}%
\providecommand \bibinfo  [0]{\@secondoftwo}%
\providecommand \bibfield  [0]{\@secondoftwo}%
\providecommand \translation [1]{[#1]}%
\providecommand \BibitemOpen [0]{}%
\providecommand \bibitemStop [0]{}%
\providecommand \bibitemNoStop [0]{.\EOS\space}%
\providecommand \EOS [0]{\spacefactor3000\relax}%
\providecommand \BibitemShut  [1]{\csname bibitem#1\endcsname}%
\let\auto@bib@innerbib\@empty
\bibitem [{\citenamefont {Holweg}\ \emph {et~al.}(1992)\citenamefont {Holweg},
  \citenamefont {Caro}, \citenamefont {Verbruggen},\ and\ \citenamefont
  {Radelaar}}]{Holweg1992}%
  \BibitemOpen
  \bibfield  {author} {\bibinfo {author} {\bibfnamefont {P.~A.~M.}\
  \bibnamefont {Holweg}}, \bibinfo {author} {\bibfnamefont {J.}~\bibnamefont
  {Caro}}, \bibinfo {author} {\bibfnamefont {A.~H.}\ \bibnamefont
  {Verbruggen}}, \ and\ \bibinfo {author} {\bibfnamefont {S.}~\bibnamefont
  {Radelaar}},\ }\bibfield  {title} {\enquote {\bibinfo {title} {Ballistic
  electron transport and two-level resistance fluctuations in noble-metal
  nanobridges},}\ }\href {\doibase 10.1103/PhysRevB.45.9311} {\bibfield
  {journal} {\bibinfo  {journal} {Phys. Rev. B}\ }\textbf {\bibinfo {volume}
  {45}},\ \bibinfo {pages} {9311} (\bibinfo {year} {1992})}\BibitemShut
  {NoStop}%
\bibitem [{\citenamefont {Galperin}\ \emph {et~al.}(1994)\citenamefont
  {Galperin}, \citenamefont {Zou},\ and\ \citenamefont {Chao}}]{Galperin1994}%
  \BibitemOpen
  \bibfield  {author} {\bibinfo {author} {\bibfnamefont {Yu.~M.}\ \bibnamefont
  {Galperin}}, \bibinfo {author} {\bibfnamefont {Nanzhi}\ \bibnamefont {Zou}},
  \ and\ \bibinfo {author} {\bibfnamefont {K.~A.}\ \bibnamefont {Chao}},\
  }\bibfield  {title} {\enquote {\bibinfo {title} {Resonant tunneling in the
  presence of a two-level fluctuator: Average transparency},}\ }\href {\doibase
  10.1103/PhysRevB.49.13728} {\bibfield  {journal} {\bibinfo  {journal} {Phys.
  Rev. B}\ }\textbf {\bibinfo {volume} {49}},\ \bibinfo {pages} {13728}
  (\bibinfo {year} {1994})}\BibitemShut {NoStop}%
\bibitem [{\citenamefont {Galperin}\ and\ \citenamefont
  {Chao}(1995)}]{Galperin1995}%
  \BibitemOpen
  \bibfield  {author} {\bibinfo {author} {\bibfnamefont {Yu.~M.}\ \bibnamefont
  {Galperin}}\ and\ \bibinfo {author} {\bibfnamefont {K.~A.}\ \bibnamefont
  {Chao}},\ }\bibfield  {title} {\enquote {\bibinfo {title} {Resonant tunneling
  in the presence of a two-level fluctuator: Low-frequency noise},}\ }\href
  {\doibase 10.1103/PhysRevB.52.12126} {\bibfield  {journal} {\bibinfo
  {journal} {Phys. Rev. B}\ }\textbf {\bibinfo {volume} {52}},\ \bibinfo
  {pages} {12126} (\bibinfo {year} {1995})}\BibitemShut {NoStop}%
\bibitem [{\citenamefont {Keijsers}\ \emph {et~al.}(1996)\citenamefont
  {Keijsers}, \citenamefont {Shklyarevskii},\ and\ \citenamefont {van
  Kempen}}]{Keijsers1996}%
  \BibitemOpen
  \bibfield  {author} {\bibinfo {author} {\bibfnamefont {R.~J.~P.}\
  \bibnamefont {Keijsers}}, \bibinfo {author} {\bibfnamefont {O.~I.}\
  \bibnamefont {Shklyarevskii}}, \ and\ \bibinfo {author} {\bibfnamefont
  {H.}~\bibnamefont {van Kempen}},\ }\bibfield  {title} {\enquote {\bibinfo
  {title} {{Point-Contact Study of Fast and Slow Two-Level Fluctuators in
  Metallic Glasses}},}\ }\href {\doibase 10.1103/PhysRevLett.77.3411}
  {\bibfield  {journal} {\bibinfo  {journal} {Phys. Rev. Lett.}\ }\textbf
  {\bibinfo {volume} {77}},\ \bibinfo {pages} {3411} (\bibinfo {year}
  {1996})}\BibitemShut {NoStop}%
\bibitem [{\citenamefont {Balkashin}\ \emph {et~al.}(1998)\citenamefont
  {Balkashin}, \citenamefont {Keijsers}, \citenamefont {van Kempen},
  \citenamefont {Kolesnichenko},\ and\ \citenamefont
  {Shklyarevskii}}]{Balkashin1998}%
  \BibitemOpen
  \bibfield  {author} {\bibinfo {author} {\bibfnamefont {O.~P.}\ \bibnamefont
  {Balkashin}}, \bibinfo {author} {\bibfnamefont {R.~J.~P.}\ \bibnamefont
  {Keijsers}}, \bibinfo {author} {\bibfnamefont {H.}~\bibnamefont {van
  Kempen}}, \bibinfo {author} {\bibfnamefont {Yu.~A.}\ \bibnamefont
  {Kolesnichenko}}, \ and\ \bibinfo {author} {\bibfnamefont {O.~I.}\
  \bibnamefont {Shklyarevskii}},\ }\bibfield  {title} {\enquote {\bibinfo
  {title} {Relaxation of two-level fluctuators in point contacts},}\ }\href
  {\doibase 10.1103/PhysRevB.58.1294} {\bibfield  {journal} {\bibinfo
  {journal} {Phys. Rev. B}\ }\textbf {\bibinfo {volume} {58}},\ \bibinfo
  {pages} {1294} (\bibinfo {year} {1998})}\BibitemShut {NoStop}%
\bibitem [{\citenamefont {Oh}\ \emph {et~al.}(2006)\citenamefont {Oh},
  \citenamefont {Cicak}, \citenamefont {Kline}, \citenamefont {Sillanp\"a\"a},
  \citenamefont {Osborn}, \citenamefont {Whittaker}, \citenamefont {Simmonds},\
  and\ \citenamefont {Pappas}}]{Oh2006}%
  \BibitemOpen
  \bibfield  {author} {\bibinfo {author} {\bibfnamefont {S.}~\bibnamefont
  {Oh}}, \bibinfo {author} {\bibfnamefont {K.}~\bibnamefont {Cicak}}, \bibinfo
  {author} {\bibfnamefont {J.~S.}\ \bibnamefont {Kline}}, \bibinfo {author}
  {\bibfnamefont {M.~A.}\ \bibnamefont {Sillanp\"a\"a}}, \bibinfo {author}
  {\bibfnamefont {K.~D.}\ \bibnamefont {Osborn}}, \bibinfo {author}
  {\bibfnamefont {J.~D.}\ \bibnamefont {Whittaker}}, \bibinfo {author}
  {\bibfnamefont {R.~W.}\ \bibnamefont {Simmonds}}, \ and\ \bibinfo {author}
  {\bibfnamefont {D.~P.}\ \bibnamefont {Pappas}},\ }\bibfield  {title}
  {\enquote {\bibinfo {title} {Elimination of two level fluctuators in
  superconducting quantum bits by an epitaxial tunnel barrier},}\ }\href
  {\doibase 10.1103/PhysRevB.74.100502} {\bibfield  {journal} {\bibinfo
  {journal} {Phys. Rev. B}\ }\textbf {\bibinfo {volume} {74}},\ \bibinfo
  {pages} {100502} (\bibinfo {year} {2006})}\BibitemShut {NoStop}%
\bibitem [{\citenamefont {Schriefl}\ \emph {et~al.}(2006)\citenamefont
  {Schriefl}, \citenamefont {Makhlin}, \citenamefont {Shnirman},\ and\
  \citenamefont {Schön}}]{Schriefl2006}%
  \BibitemOpen
  \bibfield  {author} {\bibinfo {author} {\bibfnamefont {J.}~\bibnamefont
  {Schriefl}}, \bibinfo {author} {\bibfnamefont {Y.}~\bibnamefont {Makhlin}},
  \bibinfo {author} {\bibfnamefont {A.}~\bibnamefont {Shnirman}}, \ and\
  \bibinfo {author} {\bibfnamefont {G.}~\bibnamefont {Schön}},\ }\bibfield
  {title} {\enquote {\bibinfo {title} {Decoherence from ensembles of two-level
  fluctuators},}\ }\href {\doibase 10.1088/1367-2630/8/1/001} {\bibfield
  {journal} {\bibinfo  {journal} {N. J. Phys.}\ }\textbf {\bibinfo {volume}
  {8}},\ \bibinfo {pages} {1} (\bibinfo {year} {2006})}\BibitemShut {NoStop}%
\bibitem [{\citenamefont {Zimmerman}\ \emph {et~al.}(2008)\citenamefont
  {Zimmerman}, \citenamefont {Huber}, \citenamefont {Simonds}, \citenamefont
  {Hourdakis}, \citenamefont {Fujiwara}, \citenamefont {Ono}, \citenamefont
  {Takahashi}, \citenamefont {Inokawa}, \citenamefont {Furlan},\ and\
  \citenamefont {Keller}}]{Zimmerman2008}%
  \BibitemOpen
  \bibfield  {author} {\bibinfo {author} {\bibfnamefont {N.~M.}\ \bibnamefont
  {Zimmerman}}, \bibinfo {author} {\bibfnamefont {W.~H.}\ \bibnamefont
  {Huber}}, \bibinfo {author} {\bibfnamefont {B.}~\bibnamefont {Simonds}},
  \bibinfo {author} {\bibfnamefont {E.}~\bibnamefont {Hourdakis}}, \bibinfo
  {author} {\bibfnamefont {A.}~\bibnamefont {Fujiwara}}, \bibinfo {author}
  {\bibfnamefont {Y.i}\ \bibnamefont {Ono}}, \bibinfo {author} {\bibfnamefont
  {Y.}~\bibnamefont {Takahashi}}, \bibinfo {author} {\bibfnamefont
  {H.}~\bibnamefont {Inokawa}}, \bibinfo {author} {\bibfnamefont
  {M.}~\bibnamefont {Furlan}}, \ and\ \bibinfo {author} {\bibfnamefont {M.~W.}\
  \bibnamefont {Keller}},\ }\bibfield  {title} {\enquote {\bibinfo {title}
  {{Why the long-term charge offset drift in Si single-electron tunneling
  transistors is much smaller (better) than in metal-based ones: Two-level
  fluctuator stability}},}\ }\href {https://doi.org/10.1063/1.2949700}
  {\bibfield  {journal} {\bibinfo  {journal} {J. Appl. Phys.}\ }\textbf
  {\bibinfo {volume} {104}},\ \bibinfo {pages} {033710} (\bibinfo {year}
  {2008})}\BibitemShut {NoStop}%
\bibitem [{\citenamefont {Burnett}\ \emph {et~al.}(2014)\citenamefont
  {Burnett}, \citenamefont {Faoro}, \citenamefont {Wisby}, \citenamefont
  {Gurtovoi}, \citenamefont {Chernykh}, \citenamefont {Mikhailov},
  \citenamefont {Tulin}, \citenamefont {Shaikhaidarov}, \citenamefont
  {Antonov}, \citenamefont {Meeson}, \citenamefont {Tzalenchuk},\ and\
  \citenamefont {Lindstr{\"o}m}}]{Burnett2014}%
  \BibitemOpen
  \bibfield  {author} {\bibinfo {author} {\bibfnamefont {J.}~\bibnamefont
  {Burnett}}, \bibinfo {author} {\bibfnamefont {L.}~\bibnamefont {Faoro}},
  \bibinfo {author} {\bibfnamefont {I.}~\bibnamefont {Wisby}}, \bibinfo
  {author} {\bibfnamefont {V.~L.}\ \bibnamefont {Gurtovoi}}, \bibinfo {author}
  {\bibfnamefont {A.~V.}\ \bibnamefont {Chernykh}}, \bibinfo {author}
  {\bibfnamefont {G.~M.}\ \bibnamefont {Mikhailov}}, \bibinfo {author}
  {\bibfnamefont {V.~A.}\ \bibnamefont {Tulin}}, \bibinfo {author}
  {\bibfnamefont {R.}~\bibnamefont {Shaikhaidarov}}, \bibinfo {author}
  {\bibfnamefont {V.}~\bibnamefont {Antonov}}, \bibinfo {author} {\bibfnamefont
  {P.~J.}\ \bibnamefont {Meeson}}, \bibinfo {author} {\bibfnamefont {A.~Ya}\
  \bibnamefont {Tzalenchuk}}, \ and\ \bibinfo {author} {\bibfnamefont
  {T.}~\bibnamefont {Lindstr{\"o}m}},\ }\bibfield  {title} {\enquote {\bibinfo
  {title} {Evidence for interacting two-level systems from the 1/f noise of a
  superconducting resonator},}\ }\href {https://doi.org/10.1038/ncomms5119}
  {\bibfield  {journal} {\bibinfo  {journal} {Nat. Commun.}\ }\textbf {\bibinfo
  {volume} {5}},\ \bibinfo {pages} {4119} (\bibinfo {year} {2014})},\ \bibinfo
  {note} {article}\BibitemShut {NoStop}%
\bibitem [{\citenamefont {M\"ott\"onen}\ \emph {et~al.}(2006)\citenamefont
  {M\"ott\"onen}, \citenamefont {de~Sousa}, \citenamefont {Zhang},\ and\
  \citenamefont {Whaley}}]{Mottonen2006}%
  \BibitemOpen
  \bibfield  {author} {\bibinfo {author} {\bibfnamefont {M.}~\bibnamefont
  {M\"ott\"onen}}, \bibinfo {author} {\bibfnamefont {R.}~\bibnamefont
  {de~Sousa}}, \bibinfo {author} {\bibfnamefont {J.}~\bibnamefont {Zhang}}, \
  and\ \bibinfo {author} {\bibfnamefont {K.~B.}\ \bibnamefont {Whaley}},\
  }\bibfield  {title} {\enquote {\bibinfo {title} {High-fidelity one-qubit
  operations under random telegraph noise},}\ }\href {\doibase
  10.1103/PhysRevA.73.022332} {\bibfield  {journal} {\bibinfo  {journal} {Phys.
  Rev. A}\ }\textbf {\bibinfo {volume} {73}},\ \bibinfo {pages} {022332}
  (\bibinfo {year} {2006})}\BibitemShut {NoStop}%
\bibitem [{\citenamefont {Ku}\ and\ \citenamefont {Yu}(2005)}]{Ku2005}%
  \BibitemOpen
  \bibfield  {author} {\bibinfo {author} {\bibfnamefont {L.-C.}\ \bibnamefont
  {Ku}}\ and\ \bibinfo {author} {\bibfnamefont {C.~C.}\ \bibnamefont {Yu}},\
  }\bibfield  {title} {\enquote {\bibinfo {title} {Decoherence of a {J}osephson
  qubit due to coupling to two-level systems},}\ }\href {\doibase
  10.1103/PhysRevB.72.024526} {\bibfield  {journal} {\bibinfo  {journal} {Phys.
  Rev. B}\ }\textbf {\bibinfo {volume} {72}},\ \bibinfo {pages} {024526}
  (\bibinfo {year} {2005})}\BibitemShut {NoStop}%
\bibitem [{\citenamefont {M\"uller}\ \emph {et~al.}(2009)\citenamefont
  {M\"uller}, \citenamefont {Shnirman},\ and\ \citenamefont
  {Makhlin}}]{Muller2009}%
  \BibitemOpen
  \bibfield  {author} {\bibinfo {author} {\bibfnamefont {C.}~\bibnamefont
  {M\"uller}}, \bibinfo {author} {\bibfnamefont {A.}~\bibnamefont {Shnirman}},
  \ and\ \bibinfo {author} {\bibfnamefont {Y.}~\bibnamefont {Makhlin}},\
  }\bibfield  {title} {\enquote {\bibinfo {title} {Relaxation of {J}osephson
  qubits due to strong coupling to two-level systems},}\ }\href {\doibase
  10.1103/PhysRevB.80.134517} {\bibfield  {journal} {\bibinfo  {journal} {Phys.
  Rev. B}\ }\textbf {\bibinfo {volume} {80}},\ \bibinfo {pages} {134517}
  (\bibinfo {year} {2009})}\BibitemShut {NoStop}%
\bibitem [{\citenamefont {Goetz}\ \emph {et~al.}(2016)\citenamefont {Goetz},
  \citenamefont {Deppe}, \citenamefont {Haeberlein}, \citenamefont {Wulschner},
  \citenamefont {Zollitsch}, \citenamefont {Meier}, \citenamefont {Fischer},
  \citenamefont {Eder}, \citenamefont {Xie}, \citenamefont {Fedorov},
  \citenamefont {Menzel}, \citenamefont {Marx},\ and\ \citenamefont
  {Gross}}]{Goetz2016}%
  \BibitemOpen
  \bibfield  {author} {\bibinfo {author} {\bibfnamefont {J.}~\bibnamefont
  {Goetz}}, \bibinfo {author} {\bibfnamefont {F.}~\bibnamefont {Deppe}},
  \bibinfo {author} {\bibfnamefont {M.}~\bibnamefont {Haeberlein}}, \bibinfo
  {author} {\bibfnamefont {F.}~\bibnamefont {Wulschner}}, \bibinfo {author}
  {\bibfnamefont {C.~W.}\ \bibnamefont {Zollitsch}}, \bibinfo {author}
  {\bibfnamefont {Sebastian}\ \bibnamefont {Meier}}, \bibinfo {author}
  {\bibfnamefont {M.}~\bibnamefont {Fischer}}, \bibinfo {author} {\bibfnamefont
  {P.}~\bibnamefont {Eder}}, \bibinfo {author} {\bibfnamefont {E.}~\bibnamefont
  {Xie}}, \bibinfo {author} {\bibfnamefont {K.~G.}\ \bibnamefont {Fedorov}},
  \bibinfo {author} {\bibfnamefont {E.~P.}\ \bibnamefont {Menzel}}, \bibinfo
  {author} {\bibfnamefont {A.}~\bibnamefont {Marx}}, \ and\ \bibinfo {author}
  {\bibfnamefont {R.}~\bibnamefont {Gross}},\ }\bibfield  {title} {\enquote
  {\bibinfo {title} {Loss mechanisms in superconducting thin film microwave
  resonators},}\ }\href {\doibase 10.1063/1.4939299} {\bibfield  {journal}
  {\bibinfo  {journal} {J. Appl. Phys.}\ }\textbf {\bibinfo {volume} {119}},\
  \bibinfo {pages} {015304} (\bibinfo {year} {2016})}\BibitemShut {NoStop}%
\bibitem [{\citenamefont {Giblin}\ \emph {et~al.}(2016)\citenamefont {Giblin},
  \citenamefont {See}, \citenamefont {Petrie}, \citenamefont {Janssen},
  \citenamefont {Farrer}, \citenamefont {Griffiths}, \citenamefont {Jones},
  \citenamefont {Ritchie},\ and\ \citenamefont {Kataoka}}]{Giblin2016}%
  \BibitemOpen
  \bibfield  {author} {\bibinfo {author} {\bibfnamefont {S.~P.}\ \bibnamefont
  {Giblin}}, \bibinfo {author} {\bibfnamefont {P.}~\bibnamefont {See}},
  \bibinfo {author} {\bibfnamefont {A.}~\bibnamefont {Petrie}}, \bibinfo
  {author} {\bibfnamefont {T.~J. B.~M.}\ \bibnamefont {Janssen}}, \bibinfo
  {author} {\bibfnamefont {I.}~\bibnamefont {Farrer}}, \bibinfo {author}
  {\bibfnamefont {J.~P.}\ \bibnamefont {Griffiths}}, \bibinfo {author}
  {\bibfnamefont {G.~A.~C.}\ \bibnamefont {Jones}}, \bibinfo {author}
  {\bibfnamefont {D.~A.}\ \bibnamefont {Ritchie}}, \ and\ \bibinfo {author}
  {\bibfnamefont {M.}~\bibnamefont {Kataoka}},\ }\bibfield  {title} {\enquote
  {\bibinfo {title} {High-resolution error detection in the capture process of
  a single-electron pump},}\ }\href {\doibase 10.1063/1.4939250} {\bibfield
  {journal} {\bibinfo  {journal} {Applied Physics Letters}\ }\textbf {\bibinfo
  {volume} {108}},\ \bibinfo {pages} {023502} (\bibinfo {year}
  {2016})}\BibitemShut {NoStop}%
\bibitem [{\citenamefont {Zimmerman}\ \emph {et~al.}(1997)\citenamefont
  {Zimmerman}, \citenamefont {Cobb},\ and\ \citenamefont
  {Clark}}]{Zimmerman1997}%
  \BibitemOpen
  \bibfield  {author} {\bibinfo {author} {\bibfnamefont {N.~M.}\ \bibnamefont
  {Zimmerman}}, \bibinfo {author} {\bibfnamefont {J.~L.}\ \bibnamefont {Cobb}},
  \ and\ \bibinfo {author} {\bibfnamefont {A.~F.}\ \bibnamefont {Clark}},\
  }\bibfield  {title} {\enquote {\bibinfo {title} {{Modulation of the charge of
  a single-electron transistor by distant defects}},}\ }\href {\doibase
  10.1103/PhysRevB.56.7675} {\bibfield  {journal} {\bibinfo  {journal} {Phys.
  Rev. B}\ }\textbf {\bibinfo {volume} {56}},\ \bibinfo {pages} {7675}
  (\bibinfo {year} {1997})}\BibitemShut {NoStop}%
\bibitem [{\citenamefont {Furlan}\ and\ \citenamefont
  {Lotkhov}(2003)}]{Furlan2003}%
  \BibitemOpen
  \bibfield  {author} {\bibinfo {author} {\bibfnamefont {M.}~\bibnamefont
  {Furlan}}\ and\ \bibinfo {author} {\bibfnamefont {S.~V.}\ \bibnamefont
  {Lotkhov}},\ }\bibfield  {title} {\enquote {\bibinfo {title} {Electrometry on
  charge traps with a single-electron transistor},}\ }\href {\doibase
  10.1103/PhysRevB.67.205313} {\bibfield  {journal} {\bibinfo  {journal} {Phys.
  Rev. B}\ }\textbf {\bibinfo {volume} {67}},\ \bibinfo {pages} {205313}
  (\bibinfo {year} {2003})}\BibitemShut {NoStop}%
\bibitem [{\citenamefont {Sun}\ and\ \citenamefont {Kane}(2009)}]{Sun2009}%
  \BibitemOpen
  \bibfield  {author} {\bibinfo {author} {\bibfnamefont {L.}~\bibnamefont
  {Sun}}\ and\ \bibinfo {author} {\bibfnamefont {B.~E.}\ \bibnamefont {Kane}},\
  }\bibfield  {title} {\enquote {\bibinfo {title} {Detection of a single-charge
  defect in a metal-oxide-semiconductor structure using vertically coupled {A}l
  and {S}i single-electron transistors},}\ }\href {\doibase
  10.1103/PhysRevB.80.153310} {\bibfield  {journal} {\bibinfo  {journal} {Phys.
  Rev. B}\ }\textbf {\bibinfo {volume} {80}},\ \bibinfo {pages} {153310}
  (\bibinfo {year} {2009})}\BibitemShut {NoStop}%
\bibitem [{\citenamefont {Pourkabirian}\ \emph {et~al.}(2014)\citenamefont
  {Pourkabirian}, \citenamefont {Gustafsson}, \citenamefont {Johansson},
  \citenamefont {Clarke},\ and\ \citenamefont {Delsing}}]{Pourkabirian2014}%
  \BibitemOpen
  \bibfield  {author} {\bibinfo {author} {\bibfnamefont {A.}~\bibnamefont
  {Pourkabirian}}, \bibinfo {author} {\bibfnamefont {M.~V.}\ \bibnamefont
  {Gustafsson}}, \bibinfo {author} {\bibfnamefont {G.}~\bibnamefont
  {Johansson}}, \bibinfo {author} {\bibfnamefont {J.}~\bibnamefont {Clarke}}, \
  and\ \bibinfo {author} {\bibfnamefont {P.}~\bibnamefont {Delsing}},\
  }\bibfield  {title} {\enquote {\bibinfo {title} {{Nonequilibrium Probing of
  Two-Level Charge Fluctuators Using the Step Response of a Single-Electron
  Transistor}},}\ }\href {\doibase 10.1103/PhysRevLett.113.256801} {\bibfield
  {journal} {\bibinfo  {journal} {Phys. Rev. Lett.}\ }\textbf {\bibinfo
  {volume} {113}},\ \bibinfo {pages} {256801} (\bibinfo {year}
  {2014})}\BibitemShut {NoStop}%
\bibitem [{\citenamefont {Brandes}(2008)}]{Brandes2008}%
  \BibitemOpen
  \bibfield  {author} {\bibinfo {author} {\bibfnamefont {T.}~\bibnamefont
  {Brandes}},\ }\bibfield  {title} {\enquote {\bibinfo {title} {{Waiting times
  and noise in single particle transport}},}\ }\href {\doibase
  10.1002/andp.200810306} {\bibfield  {journal} {\bibinfo  {journal} {Ann.
  Phys.}\ }\textbf {\bibinfo {volume} {17}},\ \bibinfo {pages} {477} (\bibinfo
  {year} {2008})}\BibitemShut {NoStop}%
\bibitem [{\citenamefont {Welack}\ \emph {et~al.}(2008)\citenamefont {Welack},
  \citenamefont {Esposito}, \citenamefont {Harbola},\ and\ \citenamefont
  {Mukamel}}]{Welack2008}%
  \BibitemOpen
  \bibfield  {author} {\bibinfo {author} {\bibfnamefont {S.}~\bibnamefont
  {Welack}}, \bibinfo {author} {\bibfnamefont {M.}~\bibnamefont {Esposito}},
  \bibinfo {author} {\bibfnamefont {U.}~\bibnamefont {Harbola}}, \ and\
  \bibinfo {author} {\bibfnamefont {S.}~\bibnamefont {Mukamel}},\ }\bibfield
  {title} {\enquote {\bibinfo {title} {Interference effects in the counting
  statistics of electron transfers through a double quantum dot},}\ }\href
  {\doibase 10.1103/PhysRevB.77.195315} {\bibfield  {journal} {\bibinfo
  {journal} {Phys. Rev. B}\ }\textbf {\bibinfo {volume} {77}},\ \bibinfo
  {pages} {195315} (\bibinfo {year} {2008})}\BibitemShut {NoStop}%
\bibitem [{\citenamefont {Welack}\ \emph {et~al.}(2009)\citenamefont {Welack},
  \citenamefont {Mukamel},\ and\ \citenamefont {Yan}}]{Welack2009}%
  \BibitemOpen
  \bibfield  {author} {\bibinfo {author} {\bibfnamefont {S.}~\bibnamefont
  {Welack}}, \bibinfo {author} {\bibfnamefont {S.}~\bibnamefont {Mukamel}}, \
  and\ \bibinfo {author} {\bibfnamefont {Y.~J.}\ \bibnamefont {Yan}},\
  }\bibfield  {title} {\enquote {\bibinfo {title} {Waiting time distributions
  of electron transfers through quantum dot {A}haronov-{B}ohm
  interferometers},}\ }\href {http://stacks.iop.org/0295-5075/85/i=5/a=57008}
  {\bibfield  {journal} {\bibinfo  {journal} {Europhys. Lett.}\ }\textbf
  {\bibinfo {volume} {85}},\ \bibinfo {pages} {57008} (\bibinfo {year}
  {2009})}\BibitemShut {NoStop}%
\bibitem [{\citenamefont {Welack}\ and\ \citenamefont
  {Yan}(2009)}]{Welack2009Non}%
  \BibitemOpen
  \bibfield  {author} {\bibinfo {author} {\bibfnamefont {S.}~\bibnamefont
  {Welack}}\ and\ \bibinfo {author} {\bibfnamefont {Y.~J.}\ \bibnamefont
  {Yan}},\ }\bibfield  {title} {\enquote {\bibinfo {title} {Non-{M}arkovian
  theory for the waiting time distributions of single electron transfers},}\
  }\href {\doibase 10.1063/1.3225244} {\bibfield  {journal} {\bibinfo
  {journal} {J. Chem. Phys.}\ }\textbf {\bibinfo {volume} {131}},\ \bibinfo
  {pages} {114111} (\bibinfo {year} {2009})}\BibitemShut {NoStop}%
\bibitem [{\citenamefont {Thomas}\ and\ \citenamefont
  {Flindt}(2013)}]{Thomas2013}%
  \BibitemOpen
  \bibfield  {author} {\bibinfo {author} {\bibfnamefont {K.~H.}\ \bibnamefont
  {Thomas}}\ and\ \bibinfo {author} {\bibfnamefont {C.}~\bibnamefont
  {Flindt}},\ }\bibfield  {title} {\enquote {\bibinfo {title} {Electron waiting
  times in non-{M}arkovian quantum transport},}\ }\href {\doibase
  10.1103/PhysRevB.87.121405} {\bibfield  {journal} {\bibinfo  {journal} {Phys.
  Rev. B}\ }\textbf {\bibinfo {volume} {87}},\ \bibinfo {pages} {121405}
  (\bibinfo {year} {2013})}\BibitemShut {NoStop}%
\bibitem [{\citenamefont {Tang}\ \emph {et~al.}(2014)\citenamefont {Tang},
  \citenamefont {Xu},\ and\ \citenamefont {Wang}}]{Tang2014}%
  \BibitemOpen
  \bibfield  {author} {\bibinfo {author} {\bibfnamefont {G.-M.}\ \bibnamefont
  {Tang}}, \bibinfo {author} {\bibfnamefont {F.}~\bibnamefont {Xu}}, \ and\
  \bibinfo {author} {\bibfnamefont {J.}~\bibnamefont {Wang}},\ }\bibfield
  {title} {\enquote {\bibinfo {title} {Waiting time distribution of quantum
  electronic transport in the transient regime},}\ }\href {\doibase
  10.1103/PhysRevB.89.205310} {\bibfield  {journal} {\bibinfo  {journal} {Phys.
  Rev. B}\ }\textbf {\bibinfo {volume} {89}},\ \bibinfo {pages} {205310}
  (\bibinfo {year} {2014})}\BibitemShut {NoStop}%
\bibitem [{\citenamefont {Tang}\ and\ \citenamefont
  {Wang}(2014)}]{Tang2014Full}%
  \BibitemOpen
  \bibfield  {author} {\bibinfo {author} {\bibfnamefont {G.-M.}\ \bibnamefont
  {Tang}}\ and\ \bibinfo {author} {\bibfnamefont {J.}~\bibnamefont {Wang}},\
  }\bibfield  {title} {\enquote {\bibinfo {title} {Full-counting statistics of
  charge and spin transport in the transient regime: {A} nonequilibrium
  {G}reen's function approach},}\ }\href {\doibase 10.1103/PhysRevB.90.195422}
  {\bibfield  {journal} {\bibinfo  {journal} {Phys. Rev. B}\ }\textbf {\bibinfo
  {volume} {90}},\ \bibinfo {pages} {195422} (\bibinfo {year}
  {2014})}\BibitemShut {NoStop}%
\bibitem [{\citenamefont {Sothmann}(2014)}]{Sothmann2014}%
  \BibitemOpen
  \bibfield  {author} {\bibinfo {author} {\bibfnamefont {B.}~\bibnamefont
  {Sothmann}},\ }\bibfield  {title} {\enquote {\bibinfo {title} {Electronic
  waiting-time distribution of a quantum-dot spin valve},}\ }\href {\doibase
  10.1103/PhysRevB.90.155315} {\bibfield  {journal} {\bibinfo  {journal} {Phys.
  Rev. B}\ }\textbf {\bibinfo {volume} {90}},\ \bibinfo {pages} {155315}
  (\bibinfo {year} {2014})}\BibitemShut {NoStop}%
\bibitem [{\citenamefont {Talbo}\ \emph {et~al.}(2015)\citenamefont {Talbo},
  \citenamefont {Mateos}, \citenamefont {Retailleau}, \citenamefont {Dollfus},\
  and\ \citenamefont {González}}]{Talbo2015}%
  \BibitemOpen
  \bibfield  {author} {\bibinfo {author} {\bibfnamefont {V.}~\bibnamefont
  {Talbo}}, \bibinfo {author} {\bibfnamefont {J.}~\bibnamefont {Mateos}},
  \bibinfo {author} {\bibfnamefont {S.}~\bibnamefont {Retailleau}}, \bibinfo
  {author} {\bibfnamefont {P.}~\bibnamefont {Dollfus}}, \ and\ \bibinfo
  {author} {\bibfnamefont {T.}~\bibnamefont {González}},\ }\bibfield  {title}
  {\enquote {\bibinfo {title} {Time-dependent shot noise in multi-level quantum
  dot-based single-electron devices},}\ }\href
  {http://stacks.iop.org/0268-1242/30/i=5/a=055002} {\bibfield  {journal}
  {\bibinfo  {journal} {Semicond. Sci. Technol.}\ }\textbf {\bibinfo {volume}
  {30}},\ \bibinfo {pages} {055002} (\bibinfo {year} {2015})}\BibitemShut
  {NoStop}%
\bibitem [{\citenamefont {Rudge}\ and\ \citenamefont
  {Kosov}(2016{\natexlab{a}})}]{Rudge2016}%
  \BibitemOpen
  \bibfield  {author} {\bibinfo {author} {\bibfnamefont {S.~L.}\ \bibnamefont
  {Rudge}}\ and\ \bibinfo {author} {\bibfnamefont {D.~S.}\ \bibnamefont
  {Kosov}},\ }\bibfield  {title} {\enquote {\bibinfo {title} {Distribution of
  residence times as a marker to distinguish different pathways for quantum
  transport},}\ }\href {\doibase 10.1103/PhysRevE.94.042134} {\bibfield
  {journal} {\bibinfo  {journal} {Phys. Rev. E}\ }\textbf {\bibinfo {volume}
  {94}},\ \bibinfo {pages} {042134} (\bibinfo {year}
  {2016}{\natexlab{a}})}\BibitemShut {NoStop}%
\bibitem [{\citenamefont {Rudge}\ and\ \citenamefont
  {Kosov}(2016{\natexlab{b}})}]{Rudge20162}%
  \BibitemOpen
  \bibfield  {author} {\bibinfo {author} {\bibfnamefont {S.~L.}\ \bibnamefont
  {Rudge}}\ and\ \bibinfo {author} {\bibfnamefont {D.~S.}\ \bibnamefont
  {Kosov}},\ }\bibfield  {title} {\enquote {\bibinfo {title} {Distribution of
  tunnelling times for quantum electron transport},}\ }\href {\doibase
  10.1063/1.4944493} {\bibfield  {journal} {\bibinfo  {journal} {J. Chem.
  Phys.}\ }\textbf {\bibinfo {volume} {144}},\ \bibinfo {pages} {124105}
  (\bibinfo {year} {2016}{\natexlab{b}})}\BibitemShut {NoStop}%
\bibitem [{\citenamefont {Ptaszy\ifmmode~\acute{n}\else
  \'{n}\fi{}ski}(2017)}]{Ptaszynski2017}%
  \BibitemOpen
  \bibfield  {author} {\bibinfo {author} {\bibfnamefont {K.}~\bibnamefont
  {Ptaszy\ifmmode~\acute{n}\else \'{n}\fi{}ski}},\ }\bibfield  {title}
  {\enquote {\bibinfo {title} {Nonrenewal statistics in transport through
  quantum dots},}\ }\href {\doibase 10.1103/PhysRevB.95.045306} {\bibfield
  {journal} {\bibinfo  {journal} {Phys. Rev. B}\ }\textbf {\bibinfo {volume}
  {95}},\ \bibinfo {pages} {045306} (\bibinfo {year} {2017})}\BibitemShut
  {NoStop}%
\bibitem [{\citenamefont {Rudge}\ and\ \citenamefont
  {Kosov}(2018)}]{Rudge2018}%
  \BibitemOpen
  \bibfield  {author} {\bibinfo {author} {\bibfnamefont {S.~L.}\ \bibnamefont
  {Rudge}}\ and\ \bibinfo {author} {\bibfnamefont {D.~S.}\ \bibnamefont
  {Kosov}},\ }\bibfield  {title} {\enquote {\bibinfo {title} {Distribution of
  waiting times between electron cotunneling events},}\ }\href {\doibase
  10.1103/PhysRevB.98.245402} {\bibfield  {journal} {\bibinfo  {journal} {Phys.
  Rev. B}\ }\textbf {\bibinfo {volume} {98}},\ \bibinfo {pages} {245402}
  (\bibinfo {year} {2018})}\BibitemShut {NoStop}%
\bibitem [{\citenamefont {Stegmann}\ \emph {et~al.}(2018)\citenamefont
  {Stegmann}, \citenamefont {K\"onig},\ and\ \citenamefont
  {Weiss}}]{Stegmann2018}%
  \BibitemOpen
  \bibfield  {author} {\bibinfo {author} {\bibfnamefont {P.}~\bibnamefont
  {Stegmann}}, \bibinfo {author} {\bibfnamefont {J.}~\bibnamefont {K\"onig}}, \
  and\ \bibinfo {author} {\bibfnamefont {S.}~\bibnamefont {Weiss}},\ }\bibfield
   {title} {\enquote {\bibinfo {title} {Coherent dynamics in stochastic systems
  revealed by full counting statistics},}\ }\href {\doibase
  10.1103/PhysRevB.98.035409} {\bibfield  {journal} {\bibinfo  {journal} {Phys.
  Rev. B}\ }\textbf {\bibinfo {volume} {98}},\ \bibinfo {pages} {035409}
  (\bibinfo {year} {2018})}\BibitemShut {NoStop}%
\bibitem [{\citenamefont {Kleinherbers}\ \emph {et~al.}(2018)\citenamefont
  {Kleinherbers}, \citenamefont {Stegmann},\ and\ \citenamefont
  {K\"onig}}]{Kleinherbers2018}%
  \BibitemOpen
  \bibfield  {author} {\bibinfo {author} {\bibfnamefont {E.}~\bibnamefont
  {Kleinherbers}}, \bibinfo {author} {\bibfnamefont {P}~\bibnamefont
  {Stegmann}}, \ and\ \bibinfo {author} {\bibfnamefont {P}~\bibnamefont
  {K\"onig}},\ }\bibfield  {title} {\enquote {\bibinfo {title} {Revealing
  attractive electron{\textendash}electron interaction in a quantum dot by full
  counting statistics},}\ }\href {\doibase 10.1088/1367-2630/aad14a} {\bibfield
   {journal} {\bibinfo  {journal} {New J. Phys.}\ }\textbf {\bibinfo {volume}
  {20}},\ \bibinfo {pages} {073023} (\bibinfo {year} {2018})}\BibitemShut
  {NoStop}%
\bibitem [{\citenamefont {Tang}\ \emph {et~al.}(2018)\citenamefont {Tang},
  \citenamefont {Xu}, \citenamefont {Mi},\ and\ \citenamefont
  {Wang}}]{Tang2018}%
  \BibitemOpen
  \bibfield  {author} {\bibinfo {author} {\bibfnamefont {G.}~\bibnamefont
  {Tang}}, \bibinfo {author} {\bibfnamefont {F.}~\bibnamefont {Xu}}, \bibinfo
  {author} {\bibfnamefont {S.}~\bibnamefont {Mi}}, \ and\ \bibinfo {author}
  {\bibfnamefont {J.}~\bibnamefont {Wang}},\ }\bibfield  {title} {\enquote
  {\bibinfo {title} {Spin-resolved electron waiting times in a quantum-dot spin
  valve},}\ }\href {\doibase 10.1103/PhysRevB.97.165407} {\bibfield  {journal}
  {\bibinfo  {journal} {Phys. Rev. B}\ }\textbf {\bibinfo {volume} {97}},\
  \bibinfo {pages} {165407} (\bibinfo {year} {2018})}\BibitemShut {NoStop}%
\bibitem [{\citenamefont {Rudge}\ and\ \citenamefont
  {Kosov}(2019)}]{Rudge2019}%
  \BibitemOpen
  \bibfield  {author} {\bibinfo {author} {\bibfnamefont {S.~L.}\ \bibnamefont
  {Rudge}}\ and\ \bibinfo {author} {\bibfnamefont {D.~S.}\ \bibnamefont
  {Kosov}},\ }\bibfield  {title} {\enquote {\bibinfo {title} {Nonrenewal
  statistics in quantum transport from the perspective of first-passage and
  waiting time distributions},}\ }\href {\doibase 10.1103/PhysRevB.99.115426}
  {\bibfield  {journal} {\bibinfo  {journal} {Phys. Rev. B}\ }\textbf {\bibinfo
  {volume} {99}},\ \bibinfo {pages} {115426} (\bibinfo {year}
  {2019})}\BibitemShut {NoStop}%
\bibitem [{\citenamefont {Albert}\ \emph {et~al.}(2012)\citenamefont {Albert},
  \citenamefont {Haack}, \citenamefont {Flindt},\ and\ \citenamefont
  {B\"uttiker}}]{Albert2012}%
  \BibitemOpen
  \bibfield  {author} {\bibinfo {author} {\bibfnamefont {M.}~\bibnamefont
  {Albert}}, \bibinfo {author} {\bibfnamefont {G.}~\bibnamefont {Haack}},
  \bibinfo {author} {\bibfnamefont {C.}~\bibnamefont {Flindt}}, \ and\ \bibinfo
  {author} {\bibfnamefont {M.}~\bibnamefont {B\"uttiker}},\ }\bibfield  {title}
  {\enquote {\bibinfo {title} {{E}lectron {W}aiting {T}imes in {M}esoscopic
  {C}onductors},}\ }\href {\doibase 10.1103/PhysRevLett.108.186806} {\bibfield
  {journal} {\bibinfo  {journal} {Phys. Rev. Lett.}\ }\textbf {\bibinfo
  {volume} {108}},\ \bibinfo {pages} {186806} (\bibinfo {year}
  {2012})}\BibitemShut {NoStop}%
\bibitem [{\citenamefont {Haack}\ \emph {et~al.}(2014)\citenamefont {Haack},
  \citenamefont {Albert},\ and\ \citenamefont {Flindt}}]{Haack2014}%
  \BibitemOpen
  \bibfield  {author} {\bibinfo {author} {\bibfnamefont {G.}~\bibnamefont
  {Haack}}, \bibinfo {author} {\bibfnamefont {M.}~\bibnamefont {Albert}}, \
  and\ \bibinfo {author} {\bibfnamefont {C.}~\bibnamefont {Flindt}},\
  }\bibfield  {title} {\enquote {\bibinfo {title} {Distributions of electron
  waiting times in quantum-coherent conductors},}\ }\href {\doibase
  10.1103/PhysRevB.90.205429} {\bibfield  {journal} {\bibinfo  {journal} {Phys.
  Rev. B}\ }\textbf {\bibinfo {volume} {90}},\ \bibinfo {pages} {205429}
  (\bibinfo {year} {2014})}\BibitemShut {NoStop}%
\bibitem [{\citenamefont {Seoane~Souto}\ \emph {et~al.}(2015)\citenamefont
  {Seoane~Souto}, \citenamefont {Avriller}, \citenamefont {Monreal},
  \citenamefont {Mart\'{\i}n-Rodero},\ and\ \citenamefont
  {Levy~Yeyati}}]{SeoaneSouto2015}%
  \BibitemOpen
  \bibfield  {author} {\bibinfo {author} {\bibfnamefont {R.}~\bibnamefont
  {Seoane~Souto}}, \bibinfo {author} {\bibfnamefont {R.}~\bibnamefont
  {Avriller}}, \bibinfo {author} {\bibfnamefont {R.~C.}\ \bibnamefont
  {Monreal}}, \bibinfo {author} {\bibfnamefont {A.}~\bibnamefont
  {Mart\'{\i}n-Rodero}}, \ and\ \bibinfo {author} {\bibfnamefont
  {A.}~\bibnamefont {Levy~Yeyati}},\ }\bibfield  {title} {\enquote {\bibinfo
  {title} {Transient dynamics and waiting time distribution of molecular
  junctions in the polaronic regime},}\ }\href {\doibase
  10.1103/PhysRevB.92.125435} {\bibfield  {journal} {\bibinfo  {journal} {Phys.
  Rev. B}\ }\textbf {\bibinfo {volume} {92}},\ \bibinfo {pages} {125435}
  (\bibinfo {year} {2015})}\BibitemShut {NoStop}%
\bibitem [{\citenamefont {Kosov}(2018)}]{Kosov2018}%
  \BibitemOpen
  \bibfield  {author} {\bibinfo {author} {\bibfnamefont {D.~S.}\ \bibnamefont
  {Kosov}},\ }\bibfield  {title} {\enquote {\bibinfo {title} {Waiting time
  between charging and discharging processes in molecular junctions},}\ }\href
  {\doibase 10.1063/1.5049770} {\bibfield  {journal} {\bibinfo  {journal} {J.
  Chem. Phys.}\ }\textbf {\bibinfo {volume} {149}},\ \bibinfo {pages} {164105}
  (\bibinfo {year} {2018})}\BibitemShut {NoStop}%
\bibitem [{\citenamefont {Rajabi}\ \emph {et~al.}(2013)\citenamefont {Rajabi},
  \citenamefont {P\"oltl},\ and\ \citenamefont {Governale}}]{Rajabi2013}%
  \BibitemOpen
  \bibfield  {author} {\bibinfo {author} {\bibfnamefont {L.}~\bibnamefont
  {Rajabi}}, \bibinfo {author} {\bibfnamefont {C.}~\bibnamefont {P\"oltl}}, \
  and\ \bibinfo {author} {\bibfnamefont {M.}~\bibnamefont {Governale}},\
  }\bibfield  {title} {\enquote {\bibinfo {title} {{W}aiting {T}ime
  {D}istributions for the {T}ransport through a {Q}uantum-{D}ot {T}unnel
  {C}oupled to {O}ne {N}ormal and {O}ne {S}uperconducting {L}ead},}\ }\href
  {\doibase 10.1103/PhysRevLett.111.067002} {\bibfield  {journal} {\bibinfo
  {journal} {Phys. Rev. Lett.}\ }\textbf {\bibinfo {volume} {111}},\ \bibinfo
  {pages} {067002} (\bibinfo {year} {2013})}\BibitemShut {NoStop}%
\bibitem [{\citenamefont {Dambach}\ \emph {et~al.}(2015)\citenamefont
  {Dambach}, \citenamefont {Kubala}, \citenamefont {Gramich},\ and\
  \citenamefont {Ankerhold}}]{Dambach2015}%
  \BibitemOpen
  \bibfield  {author} {\bibinfo {author} {\bibfnamefont {S.}~\bibnamefont
  {Dambach}}, \bibinfo {author} {\bibfnamefont {B.}~\bibnamefont {Kubala}},
  \bibinfo {author} {\bibfnamefont {V.}~\bibnamefont {Gramich}}, \ and\
  \bibinfo {author} {\bibfnamefont {J.}~\bibnamefont {Ankerhold}},\ }\bibfield
  {title} {\enquote {\bibinfo {title} {Time-resolved statistics of nonclassical
  light in {J}osephson photonics},}\ }\href {\doibase
  10.1103/PhysRevB.92.054508} {\bibfield  {journal} {\bibinfo  {journal} {Phys.
  Rev. B}\ }\textbf {\bibinfo {volume} {92}},\ \bibinfo {pages} {054508}
  (\bibinfo {year} {2015})}\BibitemShut {NoStop}%
\bibitem [{\citenamefont {Dambach}\ \emph {et~al.}(2016)\citenamefont
  {Dambach}, \citenamefont {Kubala},\ and\ \citenamefont
  {Ankerhold}}]{Dambach2016}%
  \BibitemOpen
  \bibfield  {author} {\bibinfo {author} {\bibfnamefont {S.}~\bibnamefont
  {Dambach}}, \bibinfo {author} {\bibfnamefont {B.}~\bibnamefont {Kubala}}, \
  and\ \bibinfo {author} {\bibfnamefont {J.}~\bibnamefont {Ankerhold}},\
  }\bibfield  {title} {\enquote {\bibinfo {title} {Time-resolved statistics of
  photon pairs in two-cavity {J}osephson photonics},}\ }\href {\doibase
  10.1002/prop.201600061} {\bibfield  {journal} {\bibinfo  {journal} {Fortschr.
  Phys.}\ ,\ \bibinfo {pages} {1}} (\bibinfo {year} {2016})}\BibitemShut
  {NoStop}%
\bibitem [{\citenamefont {Albert}\ \emph {et~al.}(2016)\citenamefont {Albert},
  \citenamefont {Chevallier},\ and\ \citenamefont {Devillard}}]{Albert2016}%
  \BibitemOpen
  \bibfield  {author} {\bibinfo {author} {\bibfnamefont {M.}~\bibnamefont
  {Albert}}, \bibinfo {author} {\bibfnamefont {D.}~\bibnamefont {Chevallier}},
  \ and\ \bibinfo {author} {\bibfnamefont {P.}~\bibnamefont {Devillard}},\
  }\bibfield  {title} {\enquote {\bibinfo {title} {Waiting times of entangled
  electrons in normal–-superconducting junctions},}\ }\href {\doibase
  http://dx.doi.org/10.1016/j.physe.2015.10.033} {\bibfield  {journal}
  {\bibinfo  {journal} {Physica E}\ }\textbf {\bibinfo {volume} {76}},\
  \bibinfo {pages} {215} (\bibinfo {year} {2016})}\BibitemShut {NoStop}%
\bibitem [{\citenamefont {Chevallier}\ \emph {et~al.}(2016)\citenamefont
  {Chevallier}, \citenamefont {Albert},\ and\ \citenamefont
  {Devillard}}]{Chevallier2016}%
  \BibitemOpen
  \bibfield  {author} {\bibinfo {author} {\bibfnamefont {D.}~\bibnamefont
  {Chevallier}}, \bibinfo {author} {\bibfnamefont {M.}~\bibnamefont {Albert}},
  \ and\ \bibinfo {author} {\bibfnamefont {P.}~\bibnamefont {Devillard}},\
  }\bibfield  {title} {\enquote {\bibinfo {title} {Probing {M}ajorana and
  {A}ndreev bound states with waiting times},}\ }\href
  {http://stacks.iop.org/0295-5075/116/i=2/a=27005} {\bibfield  {journal}
  {\bibinfo  {journal} {Europhys. Lett.}\ }\textbf {\bibinfo {volume} {116}},\
  \bibinfo {pages} {27005} (\bibinfo {year} {2016})}\BibitemShut {NoStop}%
\bibitem [{\citenamefont {Walldorf}\ \emph {et~al.}(2018)\citenamefont
  {Walldorf}, \citenamefont {Padurariu}, \citenamefont {Jauho},\ and\
  \citenamefont {Flindt}}]{Walldorf2018}%
  \BibitemOpen
  \bibfield  {author} {\bibinfo {author} {\bibfnamefont {N.}~\bibnamefont
  {Walldorf}}, \bibinfo {author} {\bibfnamefont {C.}~\bibnamefont {Padurariu}},
  \bibinfo {author} {\bibfnamefont {A.-P.}\ \bibnamefont {Jauho}}, \ and\
  \bibinfo {author} {\bibfnamefont {C.}~\bibnamefont {Flindt}},\ }\bibfield
  {title} {\enquote {\bibinfo {title} {{Electron Waiting Times of a Cooper Pair
  Splitter}},}\ }\href {\doibase 10.1103/PhysRevLett.120.087701} {\bibfield
  {journal} {\bibinfo  {journal} {Phys. Rev. Lett.}\ }\textbf {\bibinfo
  {volume} {120}},\ \bibinfo {pages} {087701} (\bibinfo {year}
  {2018})}\BibitemShut {NoStop}%
\bibitem [{\citenamefont {Mi}\ \emph {et~al.}(2018)\citenamefont {Mi},
  \citenamefont {Burset},\ and\ \citenamefont {Flindt}}]{Mi2018}%
  \BibitemOpen
  \bibfield  {author} {\bibinfo {author} {\bibfnamefont {S.}~\bibnamefont
  {Mi}}, \bibinfo {author} {\bibfnamefont {P.}~\bibnamefont {Burset}}, \ and\
  \bibinfo {author} {\bibfnamefont {C.}~\bibnamefont {Flindt}},\ }\bibfield
  {title} {\enquote {\bibinfo {title} {Electron waiting times in hybrid
  junctions with topological superconductors},}\ }\href {\doibase
  10.1038/s41598-018-34776-y} {\bibfield  {journal} {\bibinfo  {journal} {Sci.
  Rep.}\ }\textbf {\bibinfo {volume} {8}},\ \bibinfo {pages} {16828} (\bibinfo
  {year} {2018})}\BibitemShut {NoStop}%
\bibitem [{\citenamefont {Dasenbrook}\ \emph {et~al.}(2015)\citenamefont
  {Dasenbrook}, \citenamefont {Hofer},\ and\ \citenamefont
  {Flindt}}]{Dasenbrook2015}%
  \BibitemOpen
  \bibfield  {author} {\bibinfo {author} {\bibfnamefont {D.}~\bibnamefont
  {Dasenbrook}}, \bibinfo {author} {\bibfnamefont {P.~P.}\ \bibnamefont
  {Hofer}}, \ and\ \bibinfo {author} {\bibfnamefont {C.}~\bibnamefont
  {Flindt}},\ }\bibfield  {title} {\enquote {\bibinfo {title} {Electron waiting
  times in coherent conductors are correlated},}\ }\href {\doibase
  10.1103/PhysRevB.91.195420} {\bibfield  {journal} {\bibinfo  {journal} {Phys.
  Rev. B}\ }\textbf {\bibinfo {volume} {91}},\ \bibinfo {pages} {195420}
  (\bibinfo {year} {2015})}\BibitemShut {NoStop}%
\bibitem [{\citenamefont {Brange}\ \emph {et~al.}(2019)\citenamefont {Brange},
  \citenamefont {Menczel},\ and\ \citenamefont {Flindt}}]{Brange2019}%
  \BibitemOpen
  \bibfield  {author} {\bibinfo {author} {\bibfnamefont {F.}~\bibnamefont
  {Brange}}, \bibinfo {author} {\bibfnamefont {P.}~\bibnamefont {Menczel}}, \
  and\ \bibinfo {author} {\bibfnamefont {C.}~\bibnamefont {Flindt}},\
  }\bibfield  {title} {\enquote {\bibinfo {title} {Photon counting statistics
  of a microwave cavity},}\ }\href {\doibase 10.1103/PhysRevB.99.085418}
  {\bibfield  {journal} {\bibinfo  {journal} {Phys. Rev. B}\ }\textbf {\bibinfo
  {volume} {99}},\ \bibinfo {pages} {085418} (\bibinfo {year}
  {2019})}\BibitemShut {NoStop}%
\bibitem [{\citenamefont {Albert}\ \emph {et~al.}(2011)\citenamefont {Albert},
  \citenamefont {Flindt},\ and\ \citenamefont {B\"uttiker}}]{Albert2011}%
  \BibitemOpen
  \bibfield  {author} {\bibinfo {author} {\bibfnamefont {M.}~\bibnamefont
  {Albert}}, \bibinfo {author} {\bibfnamefont {C.}~\bibnamefont {Flindt}}, \
  and\ \bibinfo {author} {\bibfnamefont {M.}~\bibnamefont {B\"uttiker}},\
  }\bibfield  {title} {\enquote {\bibinfo {title} {{D}istributions of {W}aiting
  {T}imes of {D}ynamic {S}ingle-{E}lectron {E}mitters},}\ }\href {\doibase
  10.1103/PhysRevLett.107.086805} {\bibfield  {journal} {\bibinfo  {journal}
  {Phys. Rev. Lett.}\ }\textbf {\bibinfo {volume} {107}},\ \bibinfo {pages}
  {086805} (\bibinfo {year} {2011})}\BibitemShut {NoStop}%
\bibitem [{\citenamefont {Dasenbrook}\ \emph {et~al.}(2014)\citenamefont
  {Dasenbrook}, \citenamefont {Flindt},\ and\ \citenamefont
  {B\"uttiker}}]{Dasenbrook2014}%
  \BibitemOpen
  \bibfield  {author} {\bibinfo {author} {\bibfnamefont {D.}~\bibnamefont
  {Dasenbrook}}, \bibinfo {author} {\bibfnamefont {C.}~\bibnamefont {Flindt}},
  \ and\ \bibinfo {author} {\bibfnamefont {M.}~\bibnamefont {B\"uttiker}},\
  }\bibfield  {title} {\enquote {\bibinfo {title} {{F}loquet {T}heory of
  {E}lectron {W}aiting {T}imes in {Q}uantum-{C}oherent {C}onductors},}\ }\href
  {\doibase 10.1103/PhysRevLett.112.146801} {\bibfield  {journal} {\bibinfo
  {journal} {Phys. Rev. Lett.}\ }\textbf {\bibinfo {volume} {112}},\ \bibinfo
  {pages} {146801} (\bibinfo {year} {2014})}\BibitemShut {NoStop}%
\bibitem [{\citenamefont {Potanina}\ and\ \citenamefont
  {Flindt}(2017)}]{Potanina2017}%
  \BibitemOpen
  \bibfield  {author} {\bibinfo {author} {\bibfnamefont {E.}~\bibnamefont
  {Potanina}}\ and\ \bibinfo {author} {\bibfnamefont {C.}~\bibnamefont
  {Flindt}},\ }\bibfield  {title} {\enquote {\bibinfo {title} {Electron waiting
  times of a periodically driven single-electron turnstile},}\ }\href {\doibase
  10.1103/PhysRevB.96.045420} {\bibfield  {journal} {\bibinfo  {journal} {Phys.
  Rev. B}\ }\textbf {\bibinfo {volume} {96}},\ \bibinfo {pages} {045420}
  (\bibinfo {year} {2017})}\BibitemShut {NoStop}%
\bibitem [{\citenamefont {Burset}\ \emph {et~al.}(2019)\citenamefont {Burset},
  \citenamefont {Kotilahti}, \citenamefont {Moskalets},\ and\ \citenamefont
  {Flindt}}]{Burset2019}%
  \BibitemOpen
  \bibfield  {author} {\bibinfo {author} {\bibfnamefont {P.}~\bibnamefont
  {Burset}}, \bibinfo {author} {\bibfnamefont {J.}~\bibnamefont {Kotilahti}},
  \bibinfo {author} {\bibfnamefont {M.}~\bibnamefont {Moskalets}}, \ and\
  \bibinfo {author} {\bibfnamefont {C.}~\bibnamefont {Flindt}},\ }\bibfield
  {title} {\enquote {\bibinfo {title} {{Time-Domain Spectroscopy of Mesoscopic
  Conductors Using Voltage Pulses}},}\ }\href {\doibase 10.1002/qute.201970023}
  {\bibfield  {journal} {\bibinfo  {journal} {Adv. Quantum Technol.}\ }\textbf
  {\bibinfo {volume} {2}},\ \bibinfo {pages} {1970023} (\bibinfo {year}
  {2019})}\BibitemShut {NoStop}%
\bibitem [{\citenamefont {Gorman}\ \emph {et~al.}(2017)\citenamefont {Gorman},
  \citenamefont {He}, \citenamefont {House}, \citenamefont {Keizer},
  \citenamefont {Keith}, \citenamefont {Fricke}, \citenamefont {Hile},
  \citenamefont {Broome},\ and\ \citenamefont {Simmons}}]{Gorman2017}%
  \BibitemOpen
  \bibfield  {author} {\bibinfo {author} {\bibfnamefont {S.~K.}\ \bibnamefont
  {Gorman}}, \bibinfo {author} {\bibfnamefont {Y.}~\bibnamefont {He}}, \bibinfo
  {author} {\bibfnamefont {M.~G.}\ \bibnamefont {House}}, \bibinfo {author}
  {\bibfnamefont {J.~G.}\ \bibnamefont {Keizer}}, \bibinfo {author}
  {\bibfnamefont {D.}~\bibnamefont {Keith}}, \bibinfo {author} {\bibfnamefont
  {L.}~\bibnamefont {Fricke}}, \bibinfo {author} {\bibfnamefont {S.~J.}\
  \bibnamefont {Hile}}, \bibinfo {author} {\bibfnamefont {M.~A.}\ \bibnamefont
  {Broome}}, \ and\ \bibinfo {author} {\bibfnamefont {M.~Y.}\ \bibnamefont
  {Simmons}},\ }\bibfield  {title} {\enquote {\bibinfo {title} {Tunneling
  statistics for analysis of spin-readout fidelity},}\ }\href {\doibase
  10.1103/PhysRevApplied.8.034019} {\bibfield  {journal} {\bibinfo  {journal}
  {Phys. Rev. Appl.}\ }\textbf {\bibinfo {volume} {8}},\ \bibinfo {pages}
  {034019} (\bibinfo {year} {2017})}\BibitemShut {NoStop}%
\bibitem [{\citenamefont {Bagrets}\ and\ \citenamefont
  {Nazarov}(2003)}]{Bagrets2003}%
  \BibitemOpen
  \bibfield  {author} {\bibinfo {author} {\bibfnamefont {D.~A.}\ \bibnamefont
  {Bagrets}}\ and\ \bibinfo {author} {\bibfnamefont {Yu.~V.}\ \bibnamefont
  {Nazarov}},\ }\bibfield  {title} {\enquote {\bibinfo {title} {Full counting
  statistics of charge transfer in {C}oulomb blockade systems},}\ }\href
  {\doibase 10.1103/PhysRevB.67.085316} {\bibfield  {journal} {\bibinfo
  {journal} {Phys. Rev. B}\ }\textbf {\bibinfo {volume} {67}},\ \bibinfo
  {pages} {085316} (\bibinfo {year} {2003})}\BibitemShut {NoStop}%
\bibitem [{\citenamefont {Flindt}\ \emph {et~al.}(2008)\citenamefont {Flindt},
  \citenamefont {Novotn\'y}, \citenamefont {Braggio}, \citenamefont
  {Sassetti},\ and\ \citenamefont {Jauho}}]{Flindt2008}%
  \BibitemOpen
  \bibfield  {author} {\bibinfo {author} {\bibfnamefont {C.}~\bibnamefont
  {Flindt}}, \bibinfo {author} {\bibfnamefont {T.}~\bibnamefont {Novotn\'y}},
  \bibinfo {author} {\bibfnamefont {A.}~\bibnamefont {Braggio}}, \bibinfo
  {author} {\bibfnamefont {M.}~\bibnamefont {Sassetti}}, \ and\ \bibinfo
  {author} {\bibfnamefont {A.-P.}\ \bibnamefont {Jauho}},\ }\bibfield  {title}
  {\enquote {\bibinfo {title} {Counting {S}tatistics of {N}on-{M}arkovian
  {Q}uantum {S}tochastic {P}rocesses},}\ }\href {\doibase
  10.1103/PhysRevLett.100.150601} {\bibfield  {journal} {\bibinfo  {journal}
  {Phys. Rev. Lett.}\ }\textbf {\bibinfo {volume} {100}},\ \bibinfo {pages}
  {150601} (\bibinfo {year} {2008})}\BibitemShut {NoStop}%
\bibitem [{\citenamefont {Flindt}\ \emph {et~al.}(2010)\citenamefont {Flindt},
  \citenamefont {Novotn\'y}, \citenamefont {Braggio},\ and\ \citenamefont
  {Jauho}}]{Flindt2010}%
  \BibitemOpen
  \bibfield  {author} {\bibinfo {author} {\bibfnamefont {C.}~\bibnamefont
  {Flindt}}, \bibinfo {author} {\bibfnamefont {T.}~\bibnamefont {Novotn\'y}},
  \bibinfo {author} {\bibfnamefont {A.}~\bibnamefont {Braggio}}, \ and\
  \bibinfo {author} {\bibfnamefont {A.-P.}\ \bibnamefont {Jauho}},\ }\bibfield
  {title} {\enquote {\bibinfo {title} {Counting statistics of transport through
  {C}oulomb blockade nanostructures: High-order cumulants and non-{M}arkovian
  effects},}\ }\href {\doibase 10.1103/PhysRevB.82.155407} {\bibfield
  {journal} {\bibinfo  {journal} {Phys. Rev. B}\ }\textbf {\bibinfo {volume}
  {82}},\ \bibinfo {pages} {155407} (\bibinfo {year} {2010})}\BibitemShut
  {NoStop}%
\bibitem [{\citenamefont {Benito}\ \emph {et~al.}(2016)\citenamefont {Benito},
  \citenamefont {Niklas},\ and\ \citenamefont {Kohler}}]{Benito2016}%
  \BibitemOpen
  \bibfield  {author} {\bibinfo {author} {\bibfnamefont {M.}~\bibnamefont
  {Benito}}, \bibinfo {author} {\bibfnamefont {M.}~\bibnamefont {Niklas}}, \
  and\ \bibinfo {author} {\bibfnamefont {S.}~\bibnamefont {Kohler}},\
  }\bibfield  {title} {\enquote {\bibinfo {title} {Full-counting statistics of
  time-dependent conductors},}\ }\href {\doibase 10.1103/PhysRevB.94.195433}
  {\bibfield  {journal} {\bibinfo  {journal} {Phys. Rev. B}\ }\textbf {\bibinfo
  {volume} {94}},\ \bibinfo {pages} {195433} (\bibinfo {year}
  {2016})}\BibitemShut {NoStop}%
\bibitem [{\citenamefont {Talkner}\ \emph {et~al.}(2005)\citenamefont
  {Talkner}, \citenamefont {Machura}, \citenamefont {Schindler}, \citenamefont
  {H\"anggi},\ and\ \citenamefont {{\L}uczka}}]{Talkner2005}%
  \BibitemOpen
  \bibfield  {author} {\bibinfo {author} {\bibfnamefont {P.}~\bibnamefont
  {Talkner}}, \bibinfo {author} {\bibfnamefont {{\L}.}~\bibnamefont {Machura}},
  \bibinfo {author} {\bibfnamefont {M.}~\bibnamefont {Schindler}}, \bibinfo
  {author} {\bibfnamefont {P.}~\bibnamefont {H\"anggi}}, \ and\ \bibinfo
  {author} {\bibfnamefont {J.}~\bibnamefont {{\L}uczka}},\ }\bibfield  {title}
  {\enquote {\bibinfo {title} {Statistics of transition times, phase diffusion
  and synchronization in periodically driven bistable systems},}\ }\href
  {\doibase 10.1088/1367-2630/7/1/014} {\bibfield  {journal} {\bibinfo
  {journal} {New J. Phys.}\ }\textbf {\bibinfo {volume} {7}},\ \bibinfo {pages}
  {14} (\bibinfo {year} {2005})}\BibitemShut {NoStop}%
\bibitem [{\citenamefont {Naaman}\ and\ \citenamefont
  {Aumentado}(2006)}]{Naaman2006}%
  \BibitemOpen
  \bibfield  {author} {\bibinfo {author} {\bibfnamefont {O.}~\bibnamefont
  {Naaman}}\ and\ \bibinfo {author} {\bibfnamefont {J.}~\bibnamefont
  {Aumentado}},\ }\bibfield  {title} {\enquote {\bibinfo {title} {Poisson
  transition rates from time-domain measurements with a finite bandwidth},}\
  }\href {\doibase 10.1103/PhysRevLett.96.100201} {\bibfield  {journal}
  {\bibinfo  {journal} {Phys. Rev. Lett.}\ }\textbf {\bibinfo {volume} {96}},\
  \bibinfo {pages} {100201} (\bibinfo {year} {2006})}\BibitemShut {NoStop}%
\bibitem [{\citenamefont {Gustavsson}\ \emph {et~al.}(2007)\citenamefont
  {Gustavsson}, \citenamefont {Leturcq}, \citenamefont {Ihn}, \citenamefont
  {Ensslin}, \citenamefont {Reinwald},\ and\ \citenamefont
  {Wegscheider}}]{Gustavsson2007}%
  \BibitemOpen
  \bibfield  {author} {\bibinfo {author} {\bibfnamefont {S.}~\bibnamefont
  {Gustavsson}}, \bibinfo {author} {\bibfnamefont {R.}~\bibnamefont {Leturcq}},
  \bibinfo {author} {\bibfnamefont {T.}~\bibnamefont {Ihn}}, \bibinfo {author}
  {\bibfnamefont {K.}~\bibnamefont {Ensslin}}, \bibinfo {author} {\bibfnamefont
  {M.}~\bibnamefont {Reinwald}}, \ and\ \bibinfo {author} {\bibfnamefont
  {W.}~\bibnamefont {Wegscheider}},\ }\bibfield  {title} {\enquote {\bibinfo
  {title} {{Measurements of higher-order noise correlations in a quantum dot
  with a finite bandwidth detector}},}\ }\href {\doibase
  10.1103/PhysRevB.75.075314} {\bibfield  {journal} {\bibinfo  {journal} {Phys.
  Rev. B}\ }\textbf {\bibinfo {volume} {75}},\ \bibinfo {pages} {075314}
  (\bibinfo {year} {2007})}\BibitemShut {NoStop}%
\bibitem [{\citenamefont {Flindt}\ \emph {et~al.}(2009)\citenamefont {Flindt},
  \citenamefont {Fricke}, \citenamefont {Hohls}, \citenamefont {Novotny},
  \citenamefont {Netocny}, \citenamefont {Brandes},\ and\ \citenamefont
  {Haug}}]{Flindt2009}%
  \BibitemOpen
  \bibfield  {author} {\bibinfo {author} {\bibfnamefont {C.}~\bibnamefont
  {Flindt}}, \bibinfo {author} {\bibfnamefont {C.}~\bibnamefont {Fricke}},
  \bibinfo {author} {\bibfnamefont {F.}~\bibnamefont {Hohls}}, \bibinfo
  {author} {\bibfnamefont {T.}~\bibnamefont {Novotny}}, \bibinfo {author}
  {\bibfnamefont {K.}~\bibnamefont {Netocny}}, \bibinfo {author} {\bibfnamefont
  {T.}~\bibnamefont {Brandes}}, \ and\ \bibinfo {author} {\bibfnamefont
  {R.~J.}\ \bibnamefont {Haug}},\ }\bibfield  {title} {\enquote {\bibinfo
  {title} {{Universal oscillations in counting statistics}},}\ }\href {\doibase
  10.1073/pnas.0901002106} {\bibfield  {journal} {\bibinfo  {journal} {Proc.
  Natl. Acad. Sci.}\ }\textbf {\bibinfo {volume} {106}},\ \bibinfo {pages}
  {10116} (\bibinfo {year} {2009})}\BibitemShut {NoStop}%
\bibitem [{\citenamefont {Clerk}\ \emph {et~al.}(2002)\citenamefont {Clerk},
  \citenamefont {Girvin}, \citenamefont {Nguyen},\ and\ \citenamefont
  {Stone}}]{Clerk2002}%
  \BibitemOpen
  \bibfield  {author} {\bibinfo {author} {\bibfnamefont {A.~A.}\ \bibnamefont
  {Clerk}}, \bibinfo {author} {\bibfnamefont {S.~M.}\ \bibnamefont {Girvin}},
  \bibinfo {author} {\bibfnamefont {A.~K.}\ \bibnamefont {Nguyen}}, \ and\
  \bibinfo {author} {\bibfnamefont {A.~D.}\ \bibnamefont {Stone}},\ }\bibfield
  {title} {\enquote {\bibinfo {title} {{Resonant Cooper-pair tunneling: Quantum
  noise and measurement characteristics}},}\ }\href {\doibase
  10.1103/PhysRevLett.89.176804} {\bibfield  {journal} {\bibinfo  {journal}
  {Phys. Rev. Lett.}\ }\textbf {\bibinfo {volume} {89}},\ \bibinfo {pages}
  {176804} (\bibinfo {year} {2002})}\BibitemShut {NoStop}%
\bibitem [{\citenamefont {Xue}\ \emph {et~al.}(2009)\citenamefont {Xue},
  \citenamefont {Ji}, \citenamefont {Pan}, \citenamefont {Stettenheim},
  \citenamefont {Blencowe},\ and\ \citenamefont {Rimberg}}]{Xue2009}%
  \BibitemOpen
  \bibfield  {author} {\bibinfo {author} {\bibfnamefont {W.~W.}\ \bibnamefont
  {Xue}}, \bibinfo {author} {\bibfnamefont {Z.}~\bibnamefont {Ji}}, \bibinfo
  {author} {\bibfnamefont {Feng}\ \bibnamefont {Pan}}, \bibinfo {author}
  {\bibfnamefont {Joel}\ \bibnamefont {Stettenheim}}, \bibinfo {author}
  {\bibfnamefont {M.~P.}\ \bibnamefont {Blencowe}}, \ and\ \bibinfo {author}
  {\bibfnamefont {A.~J.}\ \bibnamefont {Rimberg}},\ }\bibfield  {title}
  {\enquote {\bibinfo {title} {{Measurement of quantum noise in a
  single-electron transistor near the quantum limit}},}\ }\href@noop {}
  {\bibfield  {journal} {\bibinfo  {journal} {Nat. Phys.}\ }\textbf {\bibinfo
  {volume} {5}},\ \bibinfo {pages} {660--664} (\bibinfo {year}
  {2009})}\BibitemShut {NoStop}%
\bibitem [{\citenamefont {Maisi}\ \emph {et~al.}(2014)\citenamefont {Maisi},
  \citenamefont {Kambly}, \citenamefont {Flindt},\ and\ \citenamefont
  {Pekola}}]{Maisi2014}%
  \BibitemOpen
  \bibfield  {author} {\bibinfo {author} {\bibfnamefont {V.~F.}\ \bibnamefont
  {Maisi}}, \bibinfo {author} {\bibfnamefont {D.}~\bibnamefont {Kambly}},
  \bibinfo {author} {\bibfnamefont {C.}~\bibnamefont {Flindt}}, \ and\ \bibinfo
  {author} {\bibfnamefont {J.~P.}\ \bibnamefont {Pekola}},\ }\bibfield  {title}
  {\enquote {\bibinfo {title} {{Full counting statistics of {A}ndreev
  tunneling}},}\ }\href
  {https://journals.aps.org/prl/abstract/10.1103/PhysRevLett.112.036801}
  {\bibfield  {journal} {\bibinfo  {journal} {Phys. Rev. Lett.}\ }\textbf
  {\bibinfo {volume} {112}},\ \bibinfo {pages} {036801} (\bibinfo {year}
  {2014})}\BibitemShut {NoStop}%
\end{thebibliography}%

\end{document}